\documentclass[a4paper,11pt]{article}

\usepackage{jinstpub} 

\usepackage{lineno}

\usepackage[utf8]{inputenc}
\usepackage[figuresright]{rotating}
\usepackage[section]{placeins}
\usepackage{bm} 
\usepackage{color}
\usepackage{graphicx}
\usepackage{hyperref} 
\usepackage{latexsym}
\usepackage{relsize}
\usepackage{soul}
\usepackage{subcaption}
\usepackage{xspace}

\newcounter{daggerfootnote}
\newcommand*{\daggerfootnote}[1]{%
    \setcounter{daggerfootnote}{\value{footnote}}%
    \renewcommand*{\thefootnote}{\fnsymbol{footnote}}%
    \footnote[2]{#1}%
    \setcounter{footnote}{\value{daggerfootnote}}%
    \renewcommand*{\thefootnote}{\arabic{footnote}}%
    }
    
\arxivnumber{2108.10388}

\title{\boldmath Implementation and Optimization of the PTOLEMY Transverse Drift Electromagnetic Filter}

\author[11,11a]{A.~Apponi,}
\author[10,10a]{M.G.~Betti,}
\author[7,7a]{M.~Borghesi,}
\author[19]{A.~Bosc\'a,}
\author[19]{F.~Calle,}
\author[2]{N.~Canci,}
\author[10,10a]{G.~Cavoto,}
\author[21,22]{C.~Chang,}
\author[25]{W.~Chung,\daggerfootnote{Corresponding author.}}
\author[8]{A.G.~Cocco,}
\author[14,15]{A.P.~Colijn,}
\author[2]{N.~D'Ambrosio,}
\author[16]{N.~de~Groot,}
\author[7,7a]{M.~Faverzani,}
\author[2,6] {A.~Ferella,}
\author[7]{E.~Ferri,}
\author[10,10a]{L.~Ficcadenti,}
\author[17]{P.~Garcia-Abia,}
\author[18]{G.~Garcia~Gomez-Tejedor,}
\author[12]{S.~Gariazzo,}
\author[4]{F.~Gatti,}
\author[24]{C.~Gentile,}
\author[7,7a]{A.~Giachero,}
\author[1] {Y.~Hochberg,}
\author[22,23]{Y.~Kahn,}
\author[9]{A.~Kievsky,}
\author[25]{M.~Lisanti,}
\author[8,8a]{G.~Mangano,}
\author[9,9a]{L.E.~Marcucci,}
\author[10,10a]{C.~Mariani,}
\author[19]{J.~Mart\'inez,}
\author[2]{M.~Messina,}
\author[12,13]{E.~Monticone,}
\author[7,7a]{A.~Nucciotti,}
\author[2]{D.~Orlandi,}
\author[10]{F.~Pandolfi,}
\author[2]{S.~Parlati,}
\author[19]{J.~Pedr\'os,}
\author[20]{C.~P\'erez~de~los~Heros,}
\author[8,8a]{O.~Pisanti,}
\author[10,10a]{A.D.~Polosa,}
\author[2,3]{A.~Puiu,}
\author[10,10a]{I.~Rago,}
\author[24]{Y.~Raitses,}
\author[12,13]{M.~Rajteri,}
\author[2]{N.~Rossi,}
\author[2,3]{K.~Rozwadowska,}
\author[17]{I.~Rucandio,}
\author[11,11a]{A.~Ruocco,}
\author[17]{R.~Santorelli,}
\author[5]{C.F.~Strid,}
\author[25]{A.~Tan,}
\author[25]{C.G.~Tully,}
\author[9,9a]{M.~Viviani,}
\author[16]{U.~Zeitler,}
\author[25]{F.~Zhao}

\affiliation[1]{Racah Institute of Physics, Hebrew University of Jerusalem, Jerusalem, Israel}
\affiliation[2]{INFN Laboratori Nazionali del Gran Sasso, L'Aquila, Italy}
\affiliation[3]{Gran Sasso Science Institute (GSSI), L'Aquila, Italy}
\affiliation[4]{Universit\`a di Genova e INFN Sezione di Genova, Genova, Italy }
\affiliation[5]{Johannes Gutenberg-Universität Mainz, Germany}
\affiliation[6]{Universit\`a di L'Aquila, L'Aquila, Italy}
\affiliation[7]{INFN Sezione di Milano-Bicocca, Milan, Italy }
\affiliation[7a]{Universit\`a di Milano-Bicocca, Milan, Italy}
\affiliation[8]{INFN Sezione di Napoli, Napoli, Italy}
\affiliation[8a]{Universit\`a degli Studi di Napoli Federico II, Napoli, Italy}
\affiliation[9]{INFN Sezione di Pisa, Pisa, Italy}
\affiliation[9a]{Universit\`a degli Studi di Pisa, Pisa, Italy}
\affiliation[10]{INFN Sezione di Roma 1, Roma, Italy}
\affiliation[10a]{Sapienza Universit\`a  di Roma, Roma, Italy}
\affiliation[11]{INFN Sezione di Roma 3, Roma, Italy}
\affiliation[11a]{Universit\`a  di Roma Tre, Roma, Italy}
\affiliation[12]{INFN Sezione di Torino, Torino, Italy}
\affiliation[13]{Istituto Nazionale di Ricerca Metrologica (INRiM), Torino, Italy}
\affiliation[14]{Nationaal instituut voor subatomaire fysica (NIKHEF), Amsterdam, The Netherlands}
\affiliation[15]{University of Amsterdam, Amsterdam, The Netherlands}
\affiliation[16]{Radboud University, Nijmegen, The Netherlands}
\affiliation[17]{Centro de Investigaciones Energ\'eticas, Medioambientales y Tecnol\'ogicas (CIEMAT), Madrid, Spain}
\affiliation[18]{Consejo Superior de Investigaciones Cientificas (CSIC), Madrid, Spain}
\affiliation[19]{Universidad Polit\'ecnica de Madrid, Madrid, Spain}
\affiliation[20]{Uppsala University, Uppsala, Sweden}
\affiliation[21]{Argonne National Laboratory, Chicago, IL, USA}
\affiliation[22]{Kavli Institute for Cosmological Physics, University of Chicago, Chicago, IL, USA}
\affiliation[23]{University of Illinois Urbana-Champaign, Urbana, IL, USA}
\affiliation[24]{Princeton Plasma Physics Laboratory, Princeton, NJ, USA}
\affiliation[25]{Princeton University, Princeton, NJ, USA}

\emailAdd{wonyongc@princeton.edu}

\newcommand{\CNB}{C$\nu$B\xspace}

\abstract{The PTOLEMY transverse drift filter is a new concept to enable precision analysis of the energy spectrum of electrons near the tritium $\beta$-decay endpoint. This paper details the implementation and optimization methods for successful operation of the filter. We present the first demonstrator that produces the required magnetic field properties with an iron return-flux magnet. Two methods for the setting of filter electrode voltages are detailed. The challenges of low-energy electron transport in cases of low field are discussed, such as the growth of the cyclotron radius with decreasing magnetic field, which puts a ceiling on filter performance relative to fixed filter dimensions. Additionally, low pitch angle trajectories are dominated by motion parallel to the magnetic field lines and introduce non-adiabatic conditions and curvature drift. To minimize these effects and maximize electron acceptance into the filter, we present a three-potential-well design to simultaneously drain the parallel and transverse kinetic energies throughout the length of the filter. These optimizations are shown, in simulation, to achieve low-energy electron transport from a 1\,T iron core (or 3\,T superconducting) starting field with initial kinetic energy of 18.6\,keV drained to \textless 10\,eV (\textless 1\,eV) in about 80\,cm. This result for low field operation paves the way for the first demonstrator of the PTOLEMY spectrometer for measurement of electrons near the tritium endpoint to be constructed at the Gran Sasso National Laboratary (LNGS) in Italy.}

\begin{document}
\maketitle
\flushbottom

\section{Introduction}
The precision analysis of the energy spectrum of electrons near the tritium $\beta$-decay endpoint has the potential to unlock new physics in the neutrino sector. PTOLEMY is a landmark project with the goal of being the first instrument designed to directly detect the neutrinos created in the early moments of the Big Bang, known as the Cosmic Neutrino Background (\CNB).  The concept of neutrino capture on $\beta$-decay nuclei as a detection method for the \CNB was laid out in the original paper by Steven Weinberg~\cite{Weinberg:1962zza} in 1962 and revisited by Cocco, Mangano and Messina~\cite{Cocco:2007za} in 2007 to include finite neutrino masses discovered by oscillation experiments, with tritium identified as the most promising target candidate. Further investigations on the target physics are underway~\cite{cheipesh2021heisenberg,nussinov2021quantum,kollmeier2017sdss,cheipesh2021navigating,tully2022impact,brdar2022empirical}.

Current and previous precision measurements of the tritium endpoint were based on a spectrometer technology known as the Magnetic Adiabatic Collimation and Electromagnetic (MAC-E) filter~\cite{kraus2005final,aseev2011upper,aker2019improved,aker2021first}.  An alternative concept for a high-precision compact electromagnetic filter based on transverse drift was proposed by the PTOLEMY experiment~\cite{betti2019design}. The central advantage of a transverse drift filter for the detection of the \CNB is its compact size (roughly 1\,m) which allows for the operation of many simultaneous filter elements, providing an efficient way to scale up the tritium target mass to map the neutrino sky.

The PTOLEMY experimental concept for \CNB detection was proposed in 2013~\cite{Betts:2013uya} with subsequent development of the transverse drift filter~\cite{betti2019design} and an R\&D and physics program described in~\cite{Baracchini2018, betti2019neutrino}.  The implementation details and optimization methods for successful operation of the filter are presented in this paper. A new iron-core magnet design that produces the necessary field conditions for transverse filter operation has been developed and a proof-of-concept model built at Princeton University, with the field showing good agreement with simulation. The electrode geometry and voltage configurations for the filter are detailed, including a discussion of the limiting factors that go into the selection of the filter dimensions. An iterative method and a boundary-value method are explained to demonstrate how to configure the filter voltages to account for the field transitions at the entrance and exit of the filter.

Additionally, we introduce a technique for reducing an electron's kinetic energy parallel to the magnetic field in lock step with the reduction in transverse kinetic energy.
Parallel momentum introduces additional transverse drift terms through the curvature of the magnetic field lines, causing the nominal electron trajectory to deviate from the filter midplane.
It is therefore advantageous, both for successful control of the trajectory and to increase pitch angle acceptance into the filter, to drain the parallel momentum as quickly as possible. An expanded three-well filter geometry has been developed that allows combined parallel and transverse filtering of the electron's kinetic energy.

These continued optimizations of the PTOLEMY electromagnetic filter will facilitate the construction of a compact iron-core magnetic demonstrator at the LNGS.  Previous setups of this scale have been used to set limits on relic neutrino density~\cite{robertson1991limit,LOBASHEV1999227}.  The practical implications of the filter design are important for the successful execution of the physics program with minimal systematic uncertainties from alignment and tolerances in the mechanical and electrical layouts.  The innovations on the filter design presented in this paper open up new possibilities for further reductions on the overall filter dimensions. Finally, we show that the filter can be operated ``in reverse'' as a particle accelerator rather than as a filter, opening up new potential applications for injection of charged particles into magnetic fields.
\section{Principles of Operation}

\begin{figure}[htbp]
\centering
\includegraphics[width=0.85\textwidth]{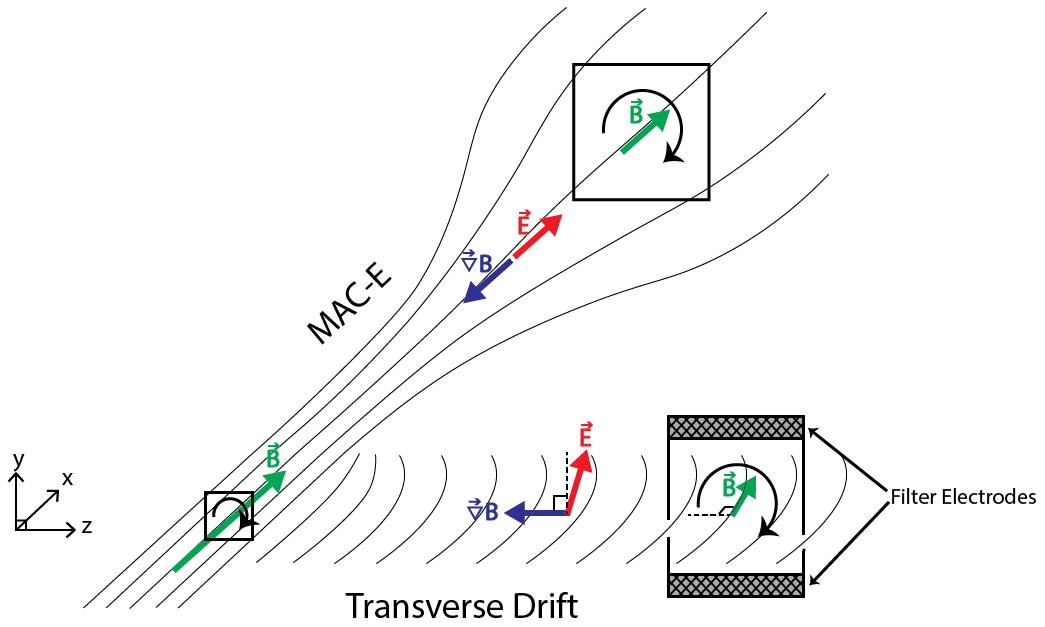}
\caption{Comparison of the magnetic field, magnetic field gradient and electric fields of the MAC-E and Transverse Drift filters.  A snapshot of the position space in the high magnetic field region of the spectrometer shows how position space expands within the aperture of the MAC-E filter as the momenta are collimated under the action of the magnetic field gradient.  For an identical snapshot in a Transverse Drift filter, a narrow window of transverse momenta remain within the aperture of the filter while all other momenta are pushed into the top or bottom electrodes.
}
\label{fig:MACEcomp}
\end{figure}

The distinguishing characteristic of a transverse drift filter in comparison with a MAC-E filter is the orientation of the magnetic field gradient with respect to the field direction~\cite{betti2019design, Beamson1980Collimating}.  The two configurations are orthogonal to one another in this respect, as shown schematically in Figure~\ref{fig:MACEcomp}.

The MAC-E filter gradient is along the field direction, hence the effect on the electron momentum is to rotate the transverse components along the field as the electron moves from a high field region to low field.  The rotation speed is conveniently computed in terms of the first adiabatic invariant, the orbital magnetic moment $\mu$, under adiabatic conditions, which are a function of the electron's parallel velocity and the magnitude of the gradient.  The electrostatic filtering is achieved with an electric field parallel to the direction of the magnetic field gradient in such a way that the electron climbs a potential barrier at a rate following the filter design where the parallel momentum component is preferably dominant to the transverse momentum component and hits zero when a turning point is reached.

In contrast, for a transverse drift filter the magnetic field gradient is orthogonal to the magnetic field direction.  The electric field is also orthogonal to the magnetic field direction, but tilted with respect to the magnetic field gradient so that net drift of the electron motion is against the direction of the magnetic field gradient.  This motion pushes the electron from a high magnetic field region to a low field.  Here again, under adiabatic conditions, the rate of reduction in the electron transverse momentum is computed using the first adiabatic invariant.  Unlike the MAC-E filter, there is no collimation effect, the electron parallel momentum is nominally unaffected by the transverse drift filter in this respect\footnote{When the electron has a non-zero parallel momentum, it is possible to reduce the parallel momentum component in lock step with the transverse momentum, as described later in this paper.}.  The work done by the magnetic field gradient term goes directly into the reduction of the electron transverse momentum.

The advantages of the transverse drift filter are in the compactness of the filter dimensions and the direct transition to a zero magnetic field region at the end of the filter.  The transverse drift filter maintains adiabatic transport for tritium endpoint electrons for over four orders of magnetic field strength within a distance of less than 1 meter.  The compact size allows for many filter elements to operate simultaneously, scaling up the effective tritium target.  The zero field region at the end of the filter is ideal for installing a transition-edge sensor (TES) microcalorimeter~\cite{rajteri2020tes}.  The microcalorimeter resolution has an energy measurement resolution goal of 0.05\,eV, providing an additional two orders of magnitude greater sensitivity to the endpoint measurement from the filter alone.
\section{Magnet Design}

In the previous work~\cite{betti2019design} which laid out the physics of the transverse drift filter, we showed that certain static but exponentially decaying configurations of electromagnetic fields can drive an electron up a potential hill while maintaining a controlled linear trajectory, and an electromagnetic filter was designed around this effect.  For a pitch 90$^\circ$ electron, the fields used, along the trajectory of the electron, were:

{\color{black}
\begin{align}
B_x & = B_0 \cos \left( \frac{x}{\lambda} \right) e^{-z/\lambda} \ , \label{eq:bxfield} \\
B_y & = 0 \ , \label{eq:byfield} \\
B_z & = - B_0 \sin \left( \frac{x}{\lambda} \right) e^{-z/\lambda} \ . \label{eq:bzfield}
\end{align}

\begin{align}
E_x & = 0 \ , \label{eq:exfield} \\
E_y & = E_0 \cos \left( \frac{y}{\lambda} \right) e^{-z/\lambda} \ , \label{eq:eyfield} \\
E_z & = - E_0 \sin \left( \frac{y}{\lambda} \right) e^{-z/\lambda} \ . \label{eq:ezfield}
\end{align}
}

The relevant drift terms, calculated in the Guiding Center System (GCS)~\cite{roederer2014particle} frame of reference, i.e. the cyclotron orbit-averaged trajectory of the electron, are the $\bm{E} \times \bm{B}$ drift, which drives transport, and the non-electric gradient-$B$ drift, which does work against the increasing potential along the trajectory. The gradient-$B$ drift is proportional to $\nabla_\perp \bm{B} / B = 1/\lambda$; this term is the radius of curvature of the field lines. It is constant if $B$ is an exponential in the transverse direction $z$ with the decay parameter $\lambda$, i.e. $B \sim e^{-z/\lambda}$. The magnitude of $\bm{E} \times \bm{B}$ drift is proportional to $E/B$, so if $\bm{E}$ is also an exponential with the same $\lambda$ as $B$, then exact canceling (in the GCS frame) is achieved between one of the $\bm{E} \times \bm{B}$ components and the gradient-$B$ drift, and the other $\bm{E} \times \bm{B}$ components are also constant. The field changes are adiabatic relative to the motion of the trajectory.

The potential hill and $\bm{E}$ are produced by electrodes lined up along the trajectory. The precise geometry of the filter electrodes is described in the next section. For $\bm{B}$ we found in~\cite{betti2019design} that solutions to Maxwell's laws in the vacuum regions between flat coils of current-carrying wire, so-called pancake coils, could satisfy the field conditions above.

For the present work we introduce an iron-core magnet design using high-permeability soft iron ($\mu/\mu_0 \approx$ 2000) that is more practical than the pancake coils for an initial field magnitude of approximately $1$\,T. Following the equivalence of $\lambda$ and the radius of curvature, we look for patterns of magnetic field lines that have an apparently constant radius of curvature along one dimension. If such a pattern is observed, it follows from the uniqueness theorem that the field magnitude in that region must be decreasing exponential with a decay parameter $\lambda$ equal to the radius of curvature of the field lines.

An iron pole-face gap magnet, such as a pair of counterposed `E'-shaped magnet cores, typically produces a region of uniform field in the air gap between the two poles. Extending transversely out from the air gap, the field decays roughly dipole in character, i.e. as $1/z^3$. This relation can be observed visually through the increasing radius of curvature of the field lines away from the gap, as all of the flux exiting one pole face must eventually return to the opposite pole face.

\begin{figure}[htbp]
\centering
\includegraphics[width=0.7\textwidth]{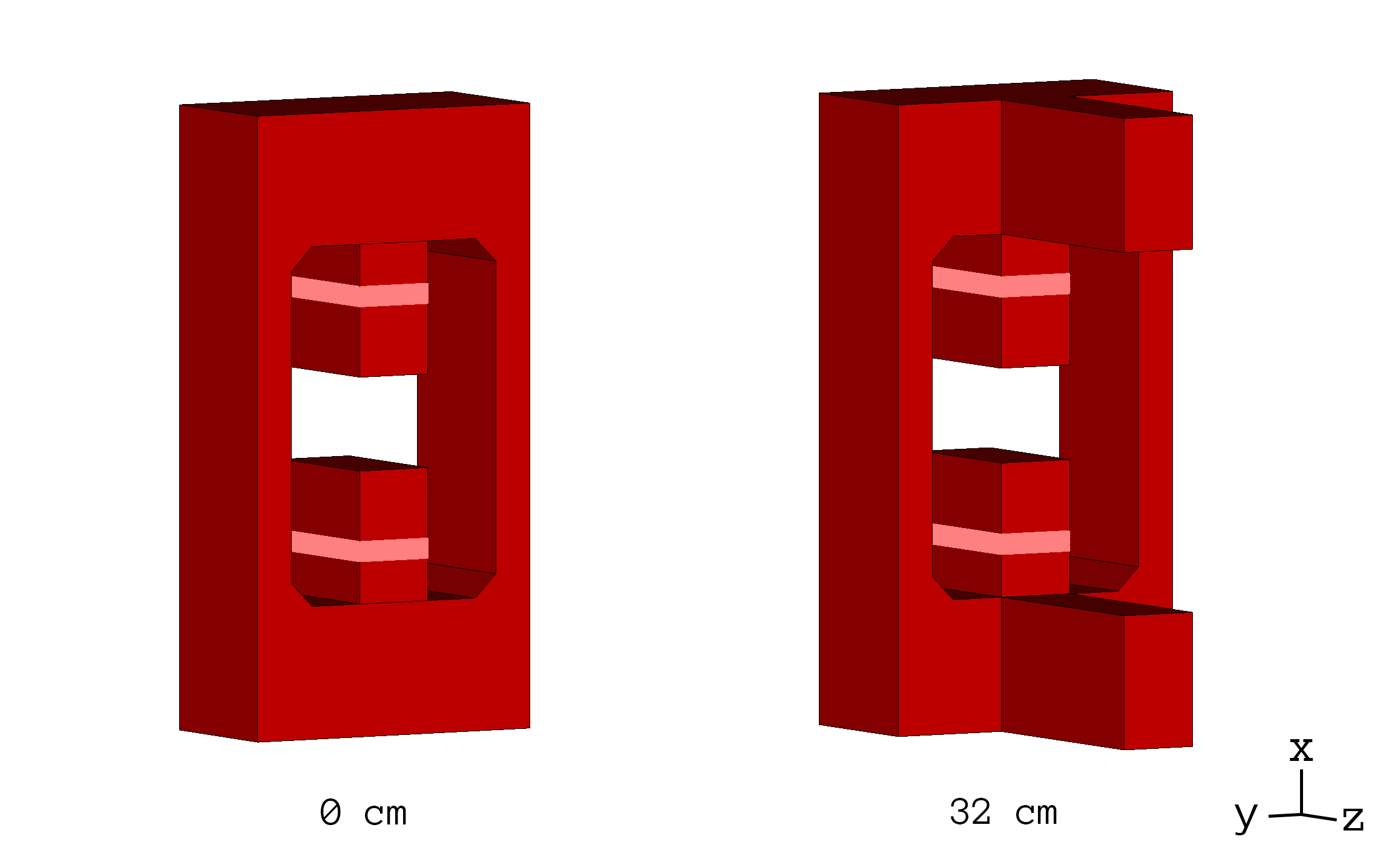}
\caption{Perspective views of a standard dual-E magnet (left) and one with 32\,cm extensions (right). Pink bands are current loops.
}
\label{fig:magnet_pers}
\end{figure}

\begin{figure}[htbp]
\centering
\includegraphics[width=1\textwidth]{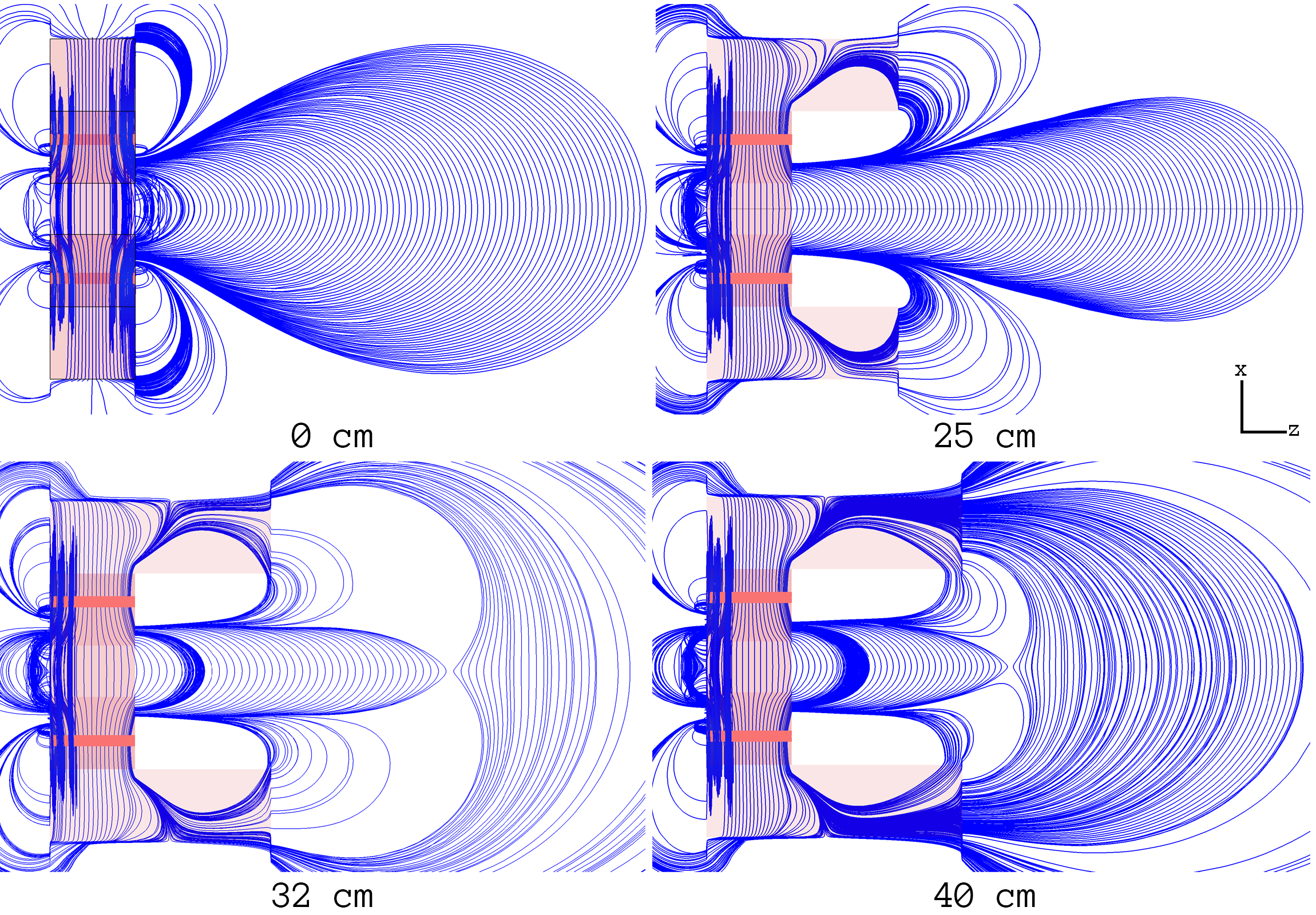}
\caption{Cross-sectional field-line views of four magnets with varying extension lengths in the plane $y=0$. The $z$ width of the standard magnet with no extensions (top left) is 20\,cm; the uniform field between the pole faces is approximately 1\,T. Extensions of length 25\,cm, 32\,cm, and 40\,cm are shown. A region of approximately constant radius of curvature is observed for an extension of length 32\,cm (bottom left).  The density of field lines is arbitrary.
}
\label{fig:magnet_fieldlines}
\end{figure}

By introducing symmetric iron extensions to the side walls of such a magnet, above and below the air gap, it is possible to turn the dipole-like field into a region of field with a constant radius of curvature. The effect of the extensions, modeled in Figure~\ref{fig:magnet_pers} as rectangular bars, is to divert some of the flux away from the return yokes and back into the vacuum above and below the transverse plane to be recycled into the air gap. The resulting flux pressure constrains the expansion of the original flux radius, and with the right extension length, a channel of field lines with apparently constant radius of curvature can be created. The effect is minimal if the horns are too short; if the extensions are too long or approach each other too closely, a flux loop between them is closed and a quadrupole point is formed in the center region where the field goes to zero and switches direction. The effect of extensions of varying length on the field profile is shown in Figure~\ref{fig:magnet_fieldlines}.

\begin{figure}[htbp]
\centering
\includegraphics[width=1\textwidth]{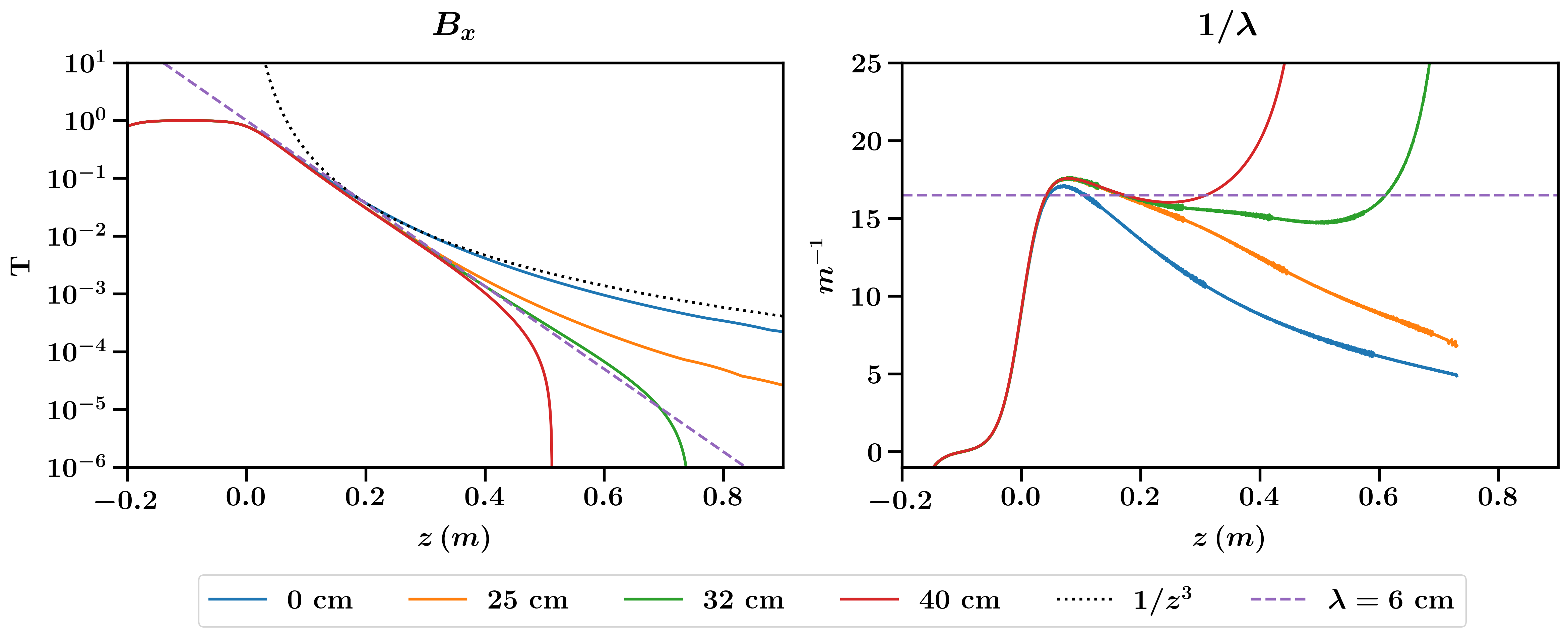}
\caption{ $B_x(z)$ and $1/\lambda (z)$ along the $z$ direction for the four magnets. Black dotted line on left chart shows the dipole-like character of the original field without extensions. The dashed line is $e^{-z/\lambda}$.
}
\label{fig:magnet_plots}
\end{figure}

The extensions need not be rectangular and fine adjustments to the location of the quadrupole point and the variance of $\lambda$ can be made by varying the shape. Such methods can be used in principle to achieve an arbitrary level of precision in $\lambda$; however, in practice, this is unnecessary as what is important is not that the field has a specific or exactly constant value of $\lambda$, but that the opposing drift components cancel out at each point in $z$ along the trajectory of the electron. This can be accomplished regardless of small deviations in $\lambda$ if the $e^{-z/\lambda}$ term of the drift components, which is used to set the voltages on the filter electrodes, is replaced by the sampled values from a precision magnetic field map of the magnet in use. Concretely, the $B_x$-component of the field is sampled along $z$ then normalized to the nearly constant $B_x$ at the air gap $z=0$ (Figure~\ref{fig:magnet_plots}):
\begin{equation}
    B_x(z)/B_x(z=0) \approx e^{-z/\lambda} \ .
\end{equation}
\section{Filter Geometry Parameters}

\begin{figure}[htbp]
\centering
\includegraphics[width=0.8\textwidth]{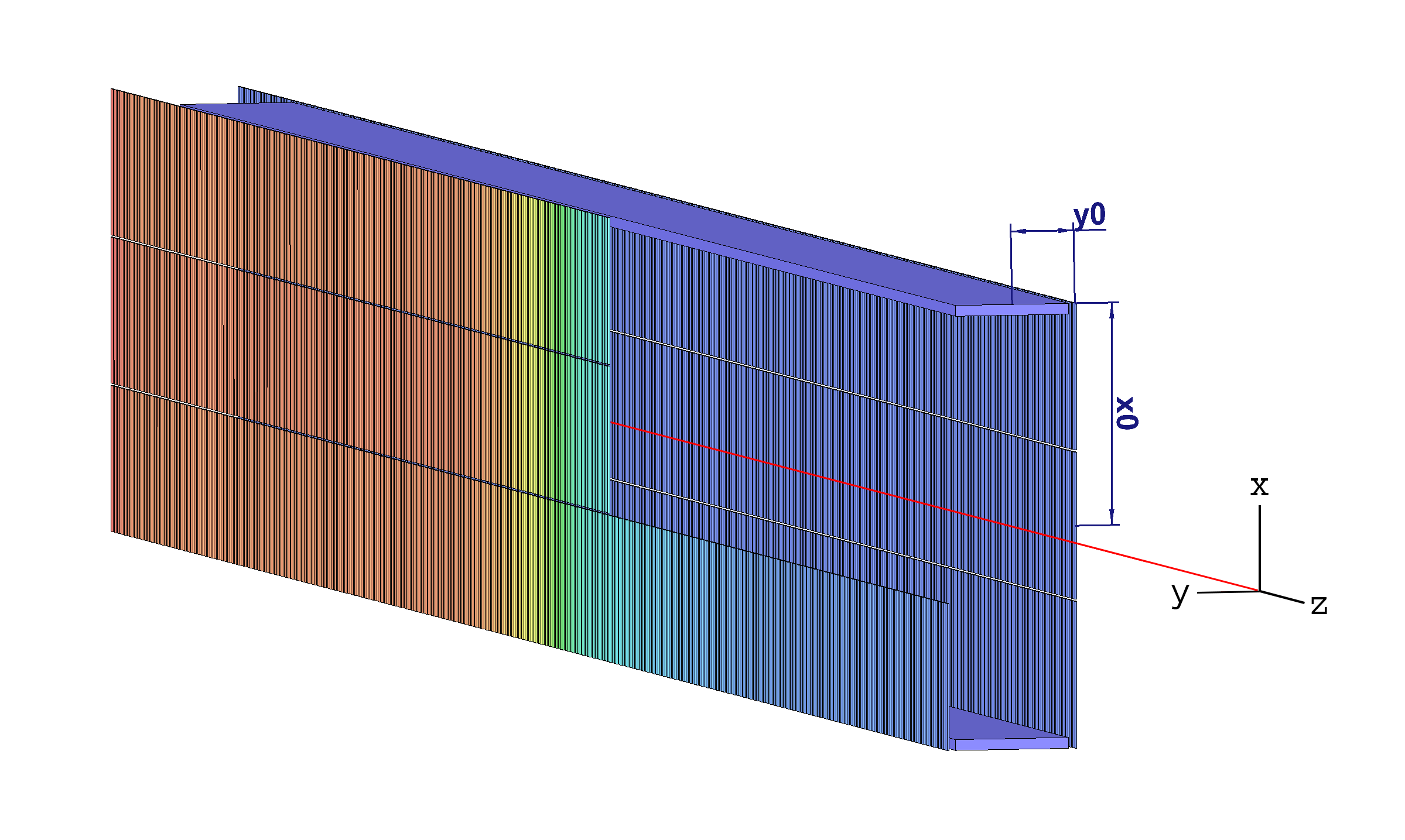}
\includegraphics[width=0.18\textwidth]{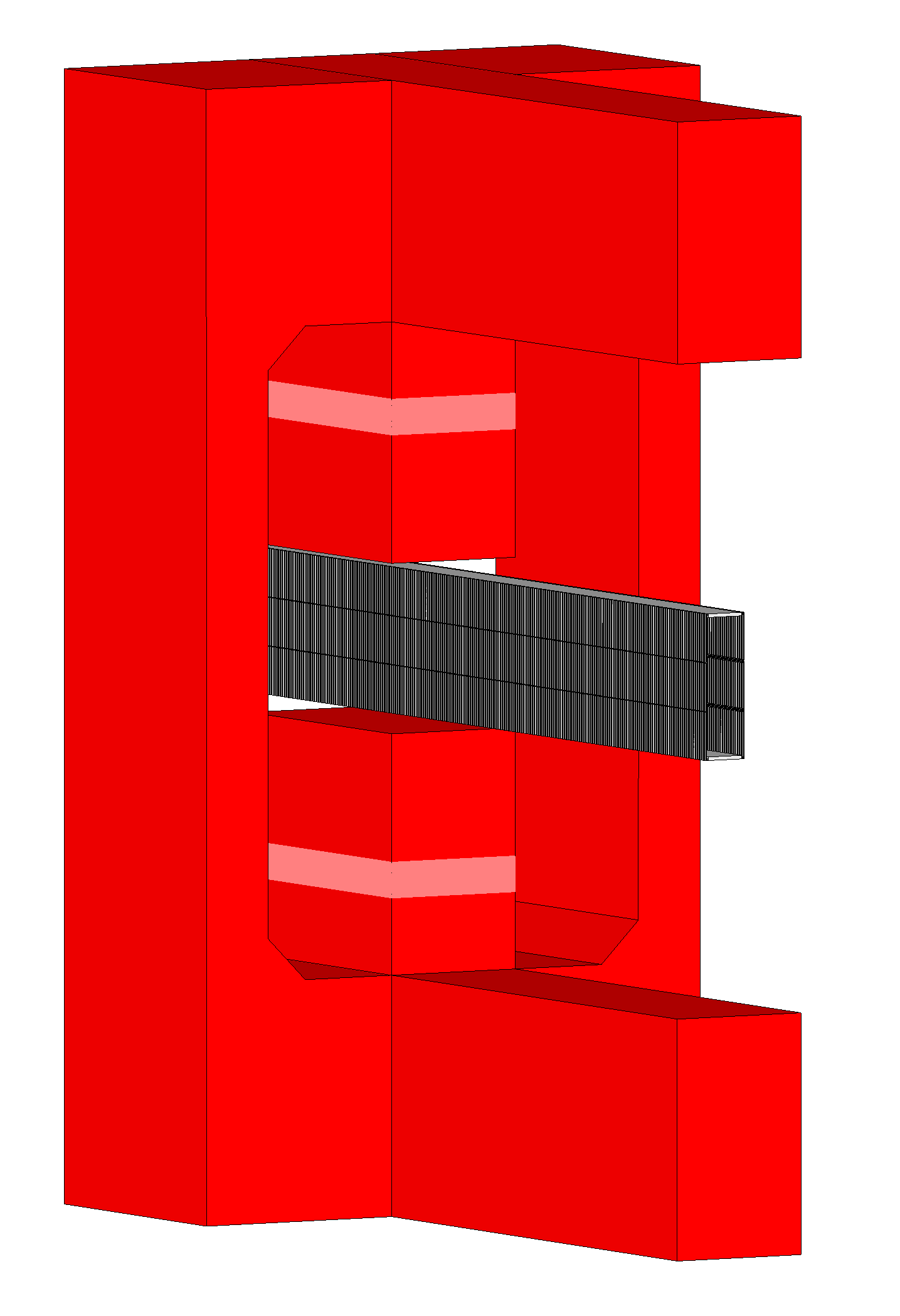}
\caption{({\it left}) Coordinate system and parameters for the filter. Part of the positive-$y$ side filter electrodes have been hidden from view. The center line of the filter, red, is the nominal GCS trajectory of the electron. Electrodes are colored by voltage. The negative-$y$ electrodes are all at constant voltage while the positive-$y$ electrodes vary. The total voltage differential along the center line is the energy drained. ({\it right insert}) The filter placed along the transverse plane of the magnet.
}
\label{fig:filter_coords}
\end{figure}

The filter geometry and coordinate system are shown in Figure~\ref{fig:filter_coords}. The magnetic field and motion of the electron parallel to it is in $x$; the transverse plane is $y-z$. The electron enters the filter through the region of uniform field in the air gap and the overall transverse drift of the trajectory is in the $+z$-direction. The electrodes responsible for the $\bm{E} \times \bm{B}$ drift are referred to as the filter electrodes and are placed a distance $\pm y_0$ from the center line ($y=0$) and span a length $2 x_0$ in $x$. Placed at $\pm x_0$ are two long electrodes that extend the full length of the filter in $z$. These are referred to as the bounce electrodes; they close off the volume enclosed by the filter electrodes and contain the electron within the filter by reflecting the parallel momentum at each end.

The parameters $x_0$ and $y_0$ are largely determined by the dimensions of the air gap of the magnet. A small aspect ratio $y_0/x_0$ is favorable to maintain field uniformity inside the filter. Nonetheless, $y_0$ should be large enough to accommodate the final cyclotron radius $\rho_c(z)$ of the electron as it grows with decreasing $B(z)$. The value of $x_0$ is best if maximized as close to the size of the air gap as possible; since the arch of the field lines is bookended by the two pole faces, it turns out $x_0 \approx \lambda$. In~\cite{betti2019design} we used a $\lambda$ of 5\,cm; the air gap in Figure~\ref{fig:magnet_fieldlines} is 12\,cm with $\lambda \approx 6$\,cm. In Figure~\ref{fig:filter_coords}, $x_0 = 5$\,cm.

To minimize $y_0$, we need the cyclotron radius $\rho_c$ as a function of $z$,

\begin{equation}
    \rho_c(z) = \frac{\sqrt{2m(T_\perp(z))}}{|q|B(z)} \
\label{eq:radius}
\end{equation}

where $q$ is the charge of the electron, $T_\perp(z) = \mu B(z)$ the transverse kinetic energy of the electron in the GCS frame, and $\mu$ the orbital magnetic moment of the electron. In terms of $B_0$, the initial field magnitude in the uniform region, and the initial cyclotron radius, $\rho_0 = \rho_c(z=0)$,

\begin{align}
    \rho_c(z)  & = \rho_0 \sqrt{B_0/B(z)} \\
                & = \rho_0 \sqrt{e^{z/\lambda}} 
\end{align}

where in the last equality we use the approximation $B(z) = B_0 e^{-z/\lambda}$. For an electron with initial $T_\perp$ = 18.6\,keV and $B_0=1$\,T, the initial radius is $\rho_0 \approx$ 0.45\,mm. With $\lambda = 6$\,cm this becomes approximately 1.25\,cm at $z = 40$\,cm, corresponding to a final kinetic energy of $\approx 20$\,eV or three orders of magnitude reduction. To accommodate a final radius of 1.25\,cm we choose $y_0 = 1.5$\,cm, which yields an aspect ratio $y_0/x_0 = 0.3$. At ratios larger than this, the field uniformity inside the filter degrades rapidly.

The ability to accommodate the growth of the cyclotron radius while maintaining field-uniformity for drift balancing is the biggest challenge against further reduction in $T_\perp$. One possibility to extend filter performance is to introduce a more elaborate filter geometry in which the aspect ratio $y_0/x_0$ is kept small while the overall dimensions increase with the radius. In general however, the additional field gradients that come with an expanding or varying geometry make this procedure complex. A more efficient way to increase filter performance is to leave the geometry intact and increase the initial $B$.

The filter power increases as $B^2$ in that a factor of three increase in $B$ results in a factor of nine decrease in the final kinetic energy. The cyclotron radius goes as $1/B$ while the energy, for a fixed radius, goes as $\rho_c^2$. If the filter dimensions are unchanged and the starting field is increased to 3\,T, the same radius 1.25\,cm is achieved at $z \approx 53$\,cm, corresponding to a final kinetic energy of $\approx 2$\,eV or four orders of magnitude reduction. The same principle of using iron extensions to redirect flux can be implemented with superconducting coils in place of an iron core to produce the necessary field.
\section{Voltage Setting Optimization}

The setting of the filter electrode voltages in an idealized, infinite-plane scenario was solved in~\cite{betti2019design}; a net $y$-drift of zero along the central line is achieved if the potential along the central line ($x$ = $y$ = 0) satisfies

\begin{align}
    \phi(z)|_{x,y=0} & = \phi_0 - \frac{\mu B_0}{|q|}\left( 1 - e^{-z/\lambda} \right) \\
    & = \phi_0 - T_\perp^0 + T_\perp^0 e^{-z/\lambda} 
\label{eq:centralphi}
\end{align}

where $\phi_0$ and $B_0$ are the initial potential and magnetic field magnitude at $z$ = 0, $T_\perp^0$ the initial transverse kinetic energy in eV, and $\mu$ is the orbital magnetic moment of the electron. To turn this into voltages for the filter electrodes, we begin with the simplifying assumption that the potential at $z$ is just the average between the two filter electrode voltages at that $z$, i.e. $\phi(z)|_{x,y=0} = [V_{y_+}(z) + V_{y_-}(z)]/2$, where $V_{y_+}(z)$, $V_{y_-}(z)$ are the voltages on the positive- and negative-$y$ side electrodes. The accuracy of this approximation increases as the aspect ratio $y_0/x_0$ decreases.

All of the negative-$y$ electrodes are set at a constant voltage which determines the total kinetic energy drained and only the positive-$y$ electrode voltages vary along $z$,

\begin{align}
    V_{y_-}(z) & = \phi_0 - \frac{\mu B_0}{|q|} \\
                & = \phi_0 - T_\perp^0 \\
    V_{y_+}(z) & = \phi_0 + \frac{\mu B_0}{|q|} \left( 2e^{-z/\lambda} -1 \right) \\
    & = \phi_0 - T_\perp^0 + 2\,T_\perp^0 e^{-z/\lambda} 
\label{eq:filtervoltages}
\end{align}

This is a schematic equation for the filter electrode voltages; as noted in the previous section, the $e^{-z/\lambda}$ term is substituted for by the normalized sampled $B_x$-component along the center line when the filter voltages are actually set.

\begin{figure}[htbp]
\centering
\includegraphics[width=1\textwidth]{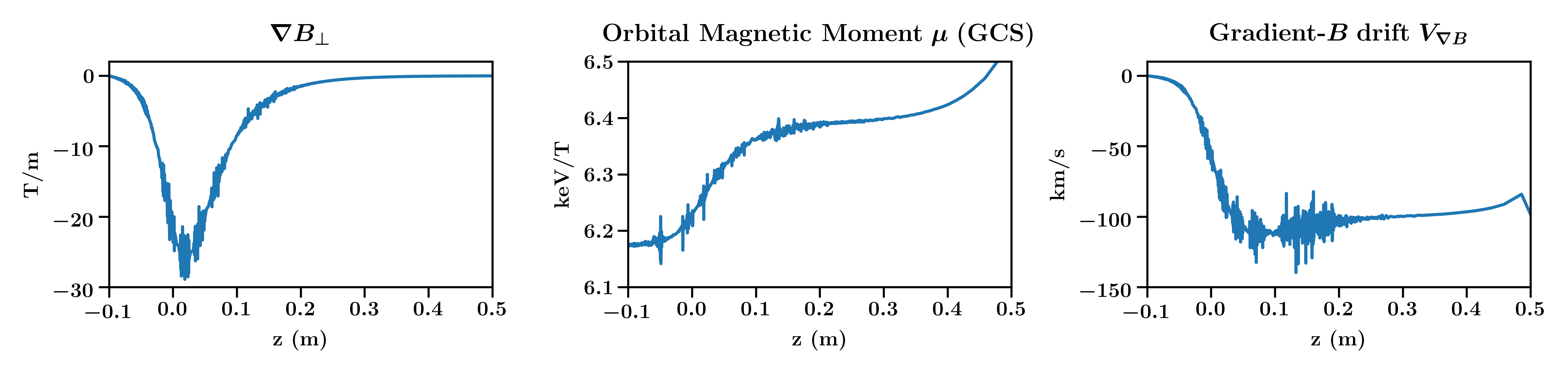}
\caption{$\nabla B_\perp$, $\mu$, and $V_{\nabla B}$ for a pitch 90$^\circ$ electron with initial transverse kinetic energy 18.6\,keV in a 3\,T initial magnetic field.
}
\label{fig:gradB_mu_VgB}
\end{figure}

Unlike the analytical conditions used in~\cite{betti2019design}, the non-zero aspect ratio of the filter and the introduction of a transition region from uniform to decaying field require corrections to the above voltages until precise drift balancing is achieved. Drift balancing can be calculated explicitly with the precision magnetic field map. The gradient-$B$ drift is nominally

\begin{equation}
    \bm{V}_{\nabla B}(z)|_{x,y=0} = - \frac{\bm{\mu} \times \bm{\nabla_\perp B(z)}}{q B(z)}
\end{equation}

where $\mu$ is taken to be adiabatically invariant in areas of low magnetic field gradient. In reality, the transition from uniform to decaying field introduces a region of high gradient at the beginning of the filter and $\mu$ is increased within the level of a few percent as shown in Figure~\ref{fig:gradB_mu_VgB}, leading to a corresponding change in gradient-$B$ drift.  Along the center line, the $B_y$ and $B_z$ components are nearly zero and therefore the magnitude $B(z) \approx B_x(z)$, and the transverse gradient $\nabla_\perp B(z) \approx d B_x/dz$, leading to

\begin{equation}
    \bm{V}_{\nabla B}(z)|_{x,y=0} = - \frac{\mu}{qB_x} \frac{d B_x}{dz} {\bm{\hat y}}
\label{eq:gradBdrift}
\end{equation}

The $y$-component of $\bm{E} \times \bm{B}$ drift that counteracts the gradient-$B$ drift is

\begin{equation}
    \bm{V}_{E \times B}^{y}(z)|_{x,y=0} = \frac{\bm{E} \times \bm{B}}{B_x^2} = \frac{E_z B_x {\bm{\hat y}}}{B_x^2} = \frac{E_z}{B_x}{\bm{\hat y}}
\label{eq:ezbxdrift}
\end{equation}

\begin{figure}[htbp]
\centering
\includegraphics[width=1\textwidth]{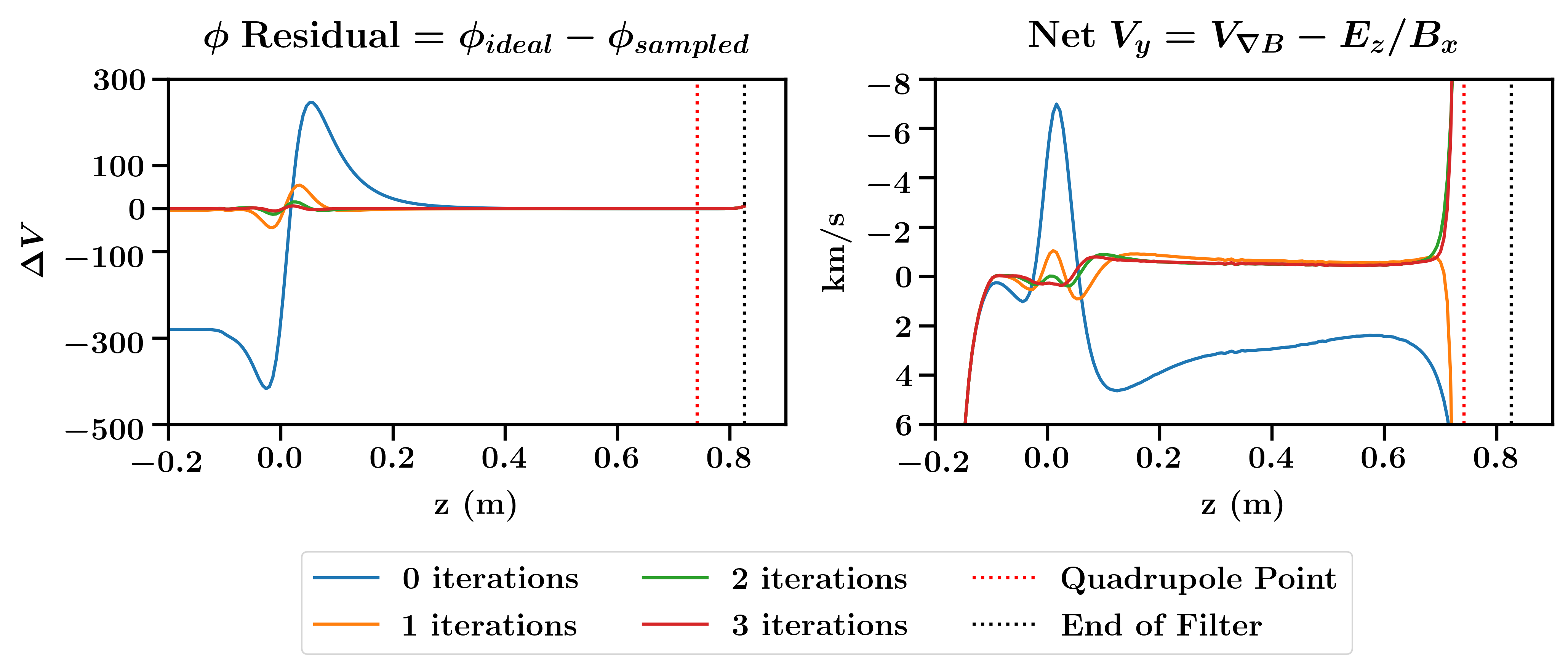}
\caption{$\Delta \phi$ and net GCS $y$-velocity along the center line for several rounds of iteration starting with a 3\,T initial field. The quadrupole point of the magnetic field and the end of the filter electrodes are indicated with dotted lines.
}
\label{fig:phidiff_netvy}
\end{figure}

The sum of the two drifts should be zero; this is the drift balancing condition and yields an expression for $E_z$ that leads to the potential~\eqref{eq:centralphi}. However, if the voltages~\eqref{eq:filtervoltages} are used as-is in the filter geometry presented in the previous section, the actual net-drift that results is non-zero (Figure~\ref{fig:phidiff_netvy}) owing to the unaccounted-for field transitions at the entrance and exit of the filter.

We present two methods to correct for these field changes. The first is a simple iterative method in which the residual between the observed potential and the idealized potential~\eqref{eq:centralphi} along the center line is added as a correction term to the voltages~\eqref{eq:filtervoltages}.  The adjustment is made only to the positive-$y$ voltages. This procedure is repeated until a desired level of convergence with~\eqref{eq:centralphi} or desired level of filter performance is achieved. Explicitly, the voltages for the $i$th iteration, $V_{y_+}[i]$, are set by

\begin{equation}
    V_{y_+}[i] = V_{y_+}[i-1] + 2 \left( \phi_{\text{ideal}} - \phi[i-1]\right)
\label{eq:correctioneq}
\end{equation}

where $V_{y_+}[i-1]$ and $\phi[i-1]$ are the voltages and potential from the previous iteration, and $\phi_{\text{ideal}}$ is the solution~\eqref{eq:centralphi}.  A factor of two is attached to the residual here to reduce the number of iterations.

\begin{figure}[htbp]
\centering
\includegraphics[width=1\textwidth]{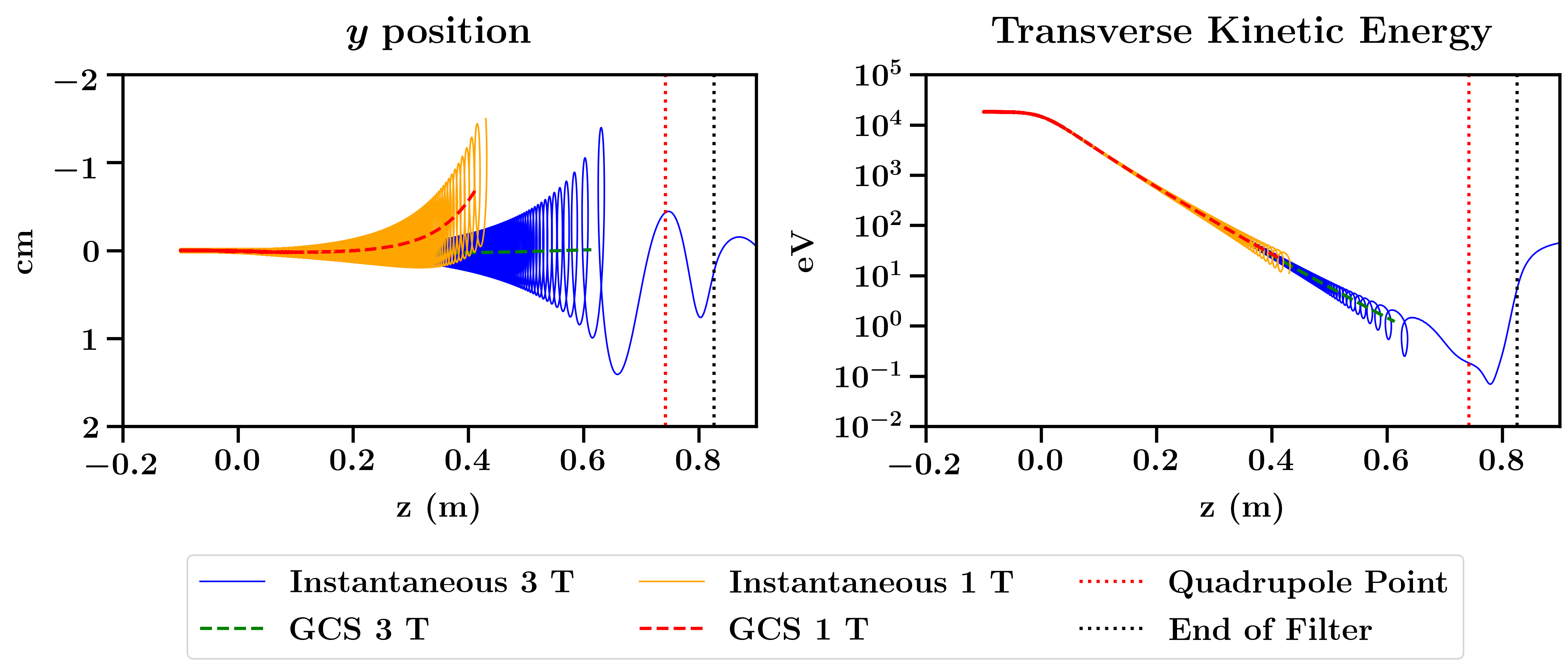}
\caption{Difference in filter performance for 1\,T vs. 3\,T starting magnetic field. Instantaneous and GCS-averaged values are shown. The initial transverse kinetic energy of the electron is 18.6\,keV. The final GCS transverse kinetic energy is 1.2\,eV for 3\,T and 9.3\,eV for 1\,T. 
The growth of the cyclotron radius of the electron as $B$ decreases puts a ceiling on filter performance for a given $y_0$. The GCS trajectories are calculated from the instantaneous trajectories by averaging values over one cyclotron orbit. The beginning and end of a single cyclotron orbit is defined by intervals in which the instantaneous $y$ and $z$ velocities of the electron change sign twice in an alternating fashion, indicating circular motion. 
}
\label{fig:y_tke}
\end{figure}

For dimensions $y_0/x_0 =1.5/5$\,cm and a starting field of 3\,T, the difference in $\phi$ compared to the ideal and the net $y$-velocity along the center line for several rounds of iteration are shown in Figure~\ref{fig:phidiff_netvy}. It is favorable to minimize the potential difference in the transition region $z = 0$ to keep the electron from falling off the center line early on; the net $y$-drift is not in practice exactly zero but is nearly constant for the majority of the filter and can be counterbalanced by offsets in the starting $y$-position of the electron. Figure~\ref{fig:y_tke} shows the transverse kinetic energy drain of the electron for starting magnetic field values of 1\,T and 3\,T, with final GCS kinetic energies of \textless 10\,eV and \textless 1\,eV respectively.

\subsection{Boundary Value Method}
\label{subsec:boundary value method}

An alternate method of solving for the filter electrode voltages based on a boundary-value method is presented here.

From (\ref{eq:gradBdrift}) and (\ref{eq:ezbxdrift}), the voltage optimization is to find the voltage on the filter electrodes such that

\begin{equation}
    E_z = - \frac{\mu}{q} \frac{d B_x}{dz}
\label{eq:y-balance}
\end{equation}

along $x,y=0$. Using the linearity of the Laplace equation, we generate the $E^{n}_{z}$, $E^{n}_{y}$, $\phi^{n}$ for the $n^{th}$ electrode with boundary value $\{V^{i}\}$ where

\begin{equation}
    V^{i} = 
    \begin{cases} 
    1 &\mbox{if } i = n ,\\
    0 & \mbox{if } i \neq n .
    \end{cases} 
\label{eq:y-balance_gen}
\end{equation}

A least-square fitting was carried out to derive the parameters $c_{n}$ such that

\begin{equation}
    \sum_{n} c_{n}\cdot E_z^{n} = - \frac{\mu}{q} \frac{d B_x}{dz}.
\label{eq:y-balance_template}
\end{equation}

The residual degrees of freedom in $c_{n}$ are constrained by matching the potential on the plane of $y=0$, $\sum_{n} c_{n}\cdot \phi^{n}$, to the kinetic energy of the electron at the entry and exit positions of the filter.

An example of the voltages on an 80-electrode demonstrator using the boundary-value method is given in section~\ref{subsec:tolerance} on tolerance estimation of the filter electrodes.

\section{Transverse Filter Speed}

The speed at which the total kinetic energy of the electron is drained by the transverse drift filter depends on the $z$-component of the $\bm{E} \times \bm{B}$ drift, which is in the positive $z$-direction,

\begin{equation}
\label{eq:vzexb}
\left. V^z_{E \times B}\right|_{x,y=0} = -(1/B) \left. \bm{\hat{y} \cdot \nabla} V(y, z) \right|_{x,y=0} = \frac{E_y}{B_x} \ \bm{\hat{z}} .
\end{equation}

It is important that the magnitude of the $z$ drift, which diverges as $1/B$, not exceed the instantaneous transverse velocity of the electron, $|v^{*}_\perp|$, until reaching the end of the filter. The GCS approximation assumes that the transverse drift is a fraction of the cyclotron velocity, giving prolate-shaped cycloid motion in the plane of the cyclotron motion. The transition to curtate-shaped cycloid motion occurs when the transverse drift velocity overtakes $|v^{*}_\perp|$, corresponding to an unraveling of the cyclotron motion in the filter frame of reference.

\begin{figure}[htbp]
\centering
\includegraphics[width=0.7\textwidth]{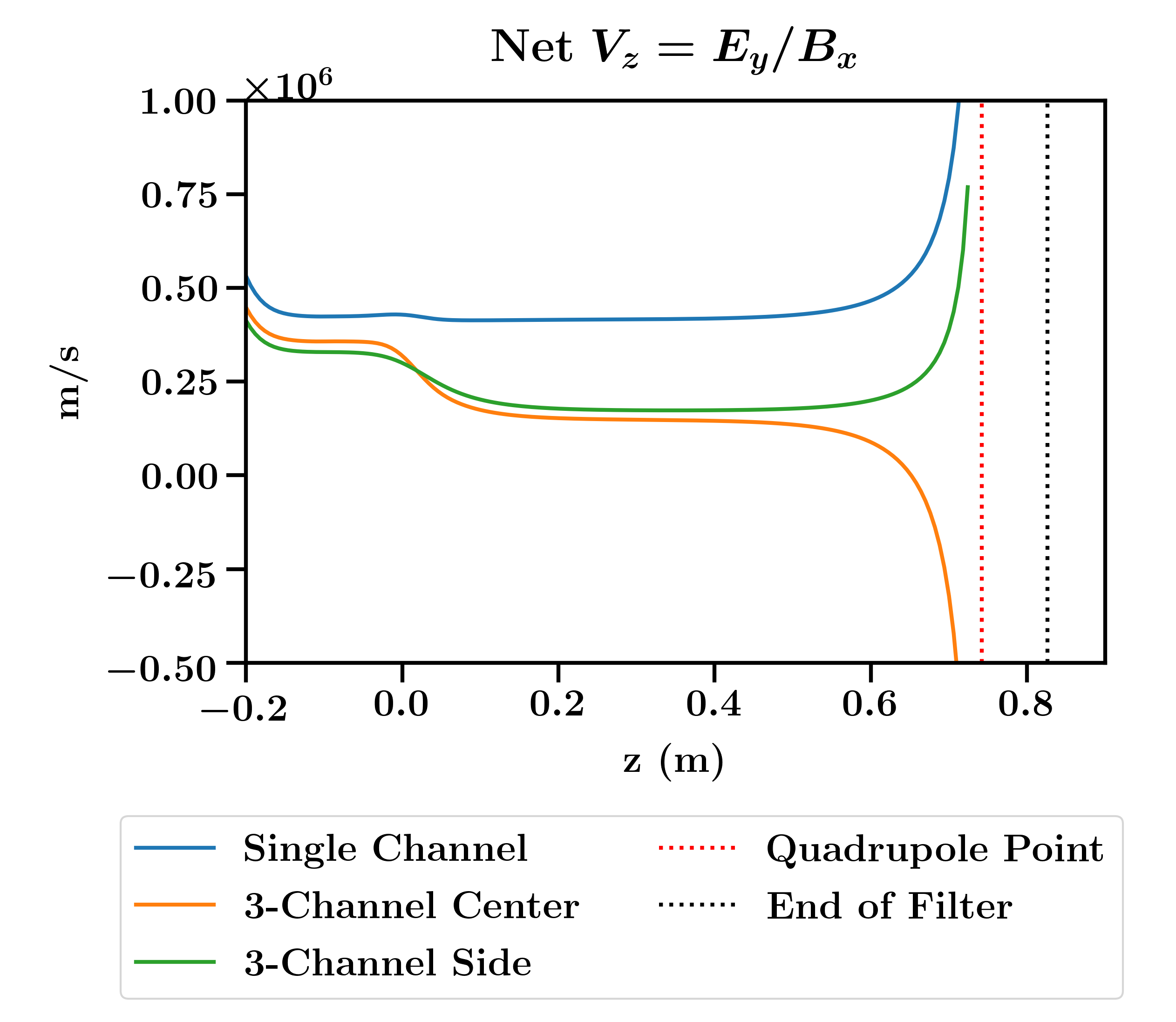}
\caption{Net $z$-velocity along the center line in the GCS frame. The asymptotic behavior is not in general physically realized as the GCS frame breaks down before this point. Nonetheless, the increase in $z$-velocity as $B$ approaches zero can be delayed by reducing the $E_y$ component of the field at the end of the filter.
}
\label{fig:netvz}
\end{figure}

\begin{figure}[htbp]
\centering
\includegraphics[width=1\textwidth]{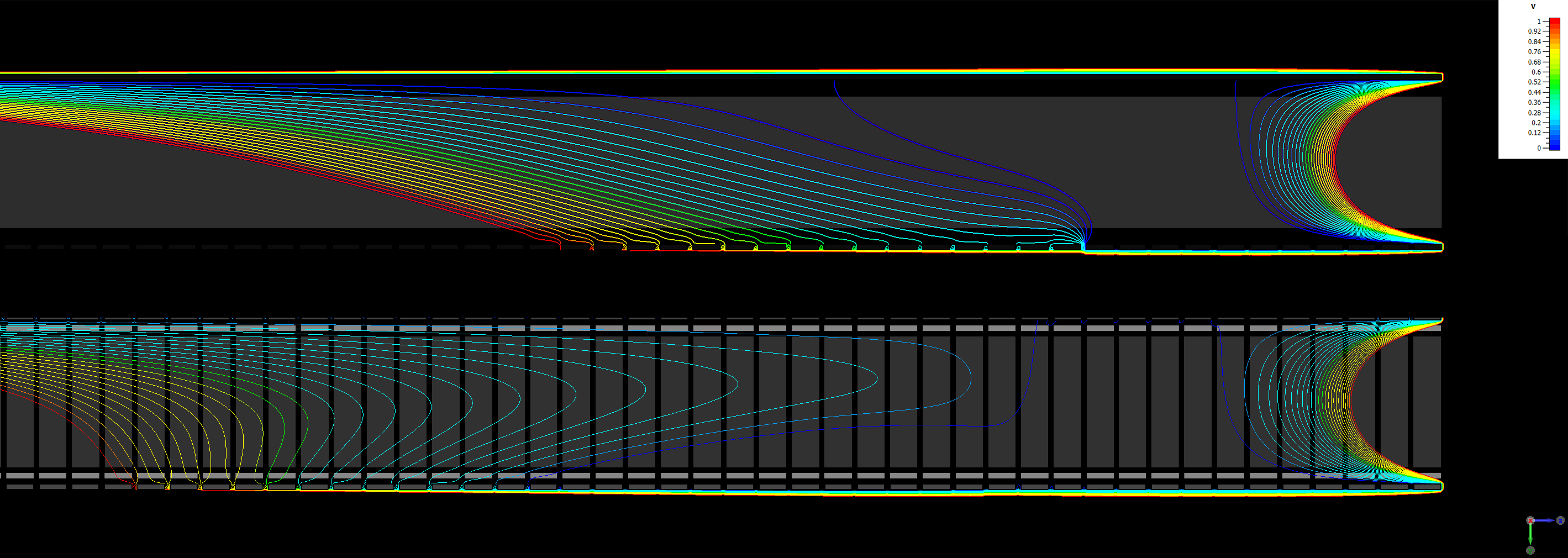}
\caption{Equipotential lines at the end of the filter in the plane $x=0$ for the single-channel design (top) and the three-channel design (bottom). In the three-channel design, the voltage settings of the side potential wells, used to drain the parallel kinetic energy, naturally form a saddle point in the center channel with a local minimum along $z$, driving the $E_y$ to zero.
}
\label{fig:eof_isolines}
\end{figure}

The divergence in $z$-velocity as $B$ decreases to zero is mitigated if the $E_y$ component of the field goes to zero faster than $B_x$ does; one way to achieve this is to form a saddle point in the potential just before the quadrupole point, with the local maximum along $x$ and the local minimum along $z$. The filter electrode voltages in this region can easily be manipulated to produce the saddle point; it also arises naturally in a three-channel filter geometry used to drain the parallel kinetic energy of the electron alongside the transverse, as shown in Figures~\ref{fig:netvz} and~\ref{fig:eof_isolines} and described in the proceeding sections.
   
\section{Kinetic Energy Parallel to the Magnetic Field}

The transverse drift filter was invented with the primary goal of draining the transverse kinetic energy of a charged particle using gradient-$B$ drift, and in the previous section the filter electrode voltages were calibrated under the assumption that all of the kinetic energy is transverse to the magnetic field, i.e. pitch 90$^\circ$. In practice the electron will also have a parallel component, and it is advantageous for successful operation of the filter for the parallel momentum to be sub-dominant to the transverse.

An electron with non-zero parallel kinetic energy inside the filter will undergo periodic `bouncing' motion in $x$; the bounce electrodes reflect electrons back to the center of the filter with a projective $\bm{E} \cdot \bm{B}$ term, which in~\cite{betti2019design} was implemented as a harmonic potential.

In general the motion of an electron inside the filter is continuous cyclotron motion with forward drift in $z$ accompanied by bouncing motion in $x$. If the parallel component becomes the dominant part of the total kinetic energy, two important effects begin to manifest. The first is the non-adiabatic nature of the trajectory where less than a single cyclotron orbit is completed before the electron has completed a single bounce, i.e. traversed the full width of the filter. The second is the transverse drift known as curvature drift.

Curvature drift originates from the centripetal forces that result as a tendency of the cyclotron motion of charged particles to follow magnetic field lines. For an exponentially falling $B$ field, the characteristic length $\lambda$ and the radius of curvature are equal, $R_c = \lambda$, and

\begin{equation}
\bm{\nabla_\perp} B =  - \frac{B}{R_c} \bm{\hat{n}}
\end{equation}

where $\bm{\hat{n}}$ is the unit vector normal to the magnetic field line curvature. In vacuum, the combined gradient-$B$ and curvature drifts are given by

\begin{equation}
\label{eq:gradbcurv}
\bm{V}_{\nabla B-C} =  \frac{1}{2}m(v_\perp^2 + 2 v_\parallel^2)  \frac{\bm{B} \times \bm{\nabla_\perp}B} {qB^3} 
 = (T_\perp + 2 T_\parallel) \frac{\bm{B} \times \bm{\nabla_\perp}B}{qB^3} 
\end{equation}

in the non-relativistic approximation~\cite{roederer2014particle}. For an equal amount of kinetic energy, the curvature drift is apparently a factor of two greater than the gradient-$B$ drift. Unlike the gradient-$B$ drift however, which is constant along the filter owing to the first adiabatic invariant $\mu$, the curvature drift depends on the instantaneous value of $v_\parallel^2$, which is therefore rapidly averaged over successive bounces in the filter to have an effective bounced-averaged value $\langle v_\parallel^2 \rangle$. For a filter geometry symmetric in $x$ the maximum value $|v_\parallel^{max}|$ is attained at $x=0$, and for a harmonic bounce potential along the $B$ field direction, the average value of $\langle v_\parallel^2 \rangle = (v_\parallel^{max})^2/2$. Plugging this back into~\eqref{eq:gradbcurv}, the factor of two relative to the gradient-$B$ drift disappears. 

Therefore, the total gradient-$B$ and curvature transverse drift is proportional to the total kinetic energy $T(z)$ of the electron at $x=0$ in the filter and points in the negative $y$-direction,
\begin{equation}
\label{eq:kedrain}
\bm{V}_{\nabla B-C} \sim \left. -\frac{T(z)}{B(z) R_c} \right|_{x=0}
\bm{\hat{y}} \ .
\end{equation}


Additionally, as described in~\cite{roederer2014particle} in the discussion following Eq.~(2.13) on particle inertia, the component of transverse drift along the normal of the $B$ field, i.e. along $z$, introduces an additional contribution to the curvature drift that scales as $|v_\parallel V^z_{E \times B}|$ rather than $v_\parallel^2$.  Therefore, at the end of the filter, the effects of curvature drift can be dramatic if $V^z_{E \times B}$ is a substantial fraction of $|v^{*}_\perp|$. The $V^z_{E \times B}$ drift is mitigated by use of a saddle point, nonetheless it is advantageous to also drain the parallel energy faster relative to the transverse to avoid this runaway drift.

\section{Bounce Electrodes, Field Wires and Side-Well Potentials}

The parallel kinetic energy of an electron inside the filter can be drained by modifying the design into a three-channel geometry as follows. Consider an electron undergoing bounce motion in the filter. At the turning points of the motion, the parallel kinetic energy goes to zero while the transverse kinetic energy is largely unchanged. When the electron is reflected back towards the center of the filter by the bounce electrodes, the parallel kinetic energy returns to its maximum value at $x=0$. However, because the electron is also moving forward in $z$ as the bounce occurs, the potential at $x=0$ upon return to the center is not the same as it was when it left; it has increased. In the GCS frame, since this increase in potential is along the direction of the bounce motion, and not in the transverse, the parallel kinetic energy is the one that is drained, in an amount nominally equal to the potential difference at $x=0$ before and after the bounce.

\begin{figure}[htbp]
\centering
\includegraphics[width=1\textwidth]{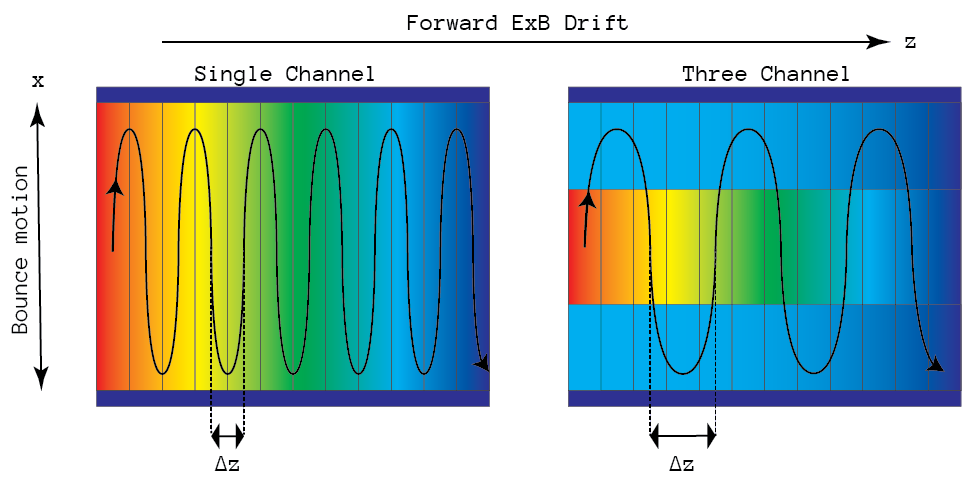}
\caption{Schematic drawing of parallel momentum draining in the single channel vs. three channel filter designs. Color indicates potential, from red (lowest) to blue (highest). Dark blue strips at the top and bottom are bounce electrodes, with filter electrodes drawn as vertical bars along the trajectory. In the three channel design, the side channels are set at a higher voltage than the center, approximately equal to the parallel kinetic energy along the trajectory. The electron lingers in the side well before returning to the center, thus re-entering at a higher potential and draining more parallel energy than it would have in the single channel design.
}
\label{fig:parallel_drain_schematic}
\end{figure}

This effect can be observed in the original single-channel filter, but the parallel drain is small since the forward displacement in $z$ between bounces is small and the potential gradient is not in general oriented in the bounce direction. The amount of parallel drain can be increased if the electron is made to linger in the side region of the filter before returning to the center, increasing the forward displacement in $z$. A schematic drawing of this mechanism is shown in Figure~\ref{fig:parallel_drain_schematic}.

This can be achieved to good precision if the initial parallel kinetic energy of the electron is known. The interior of the filter is split into three discrete potential wells; we call them the center- and side-wells. To the center potential, previously configured only for the transverse kinetic energy drain, is now added the parallel draining term. The center potential at the end of the filter remains the total energy drained. The side potential at $z$ is increased (decreased, as it were, for an electron) by an amount equal to the parallel kinetic energy of the electron at $z$. Since the potential step between side-center-side is in the direction of parallel motion, when the electron enters the side-well from the center its parallel kinetic energy is reduced to nearly zero for an extended period before bouncing back, compared to the single-well design where the bounce is nearly instantaneous, thus causing it to linger there while undergoing transverse drain forward in $z$, before eventually bouncing back to the center as before, achieving the desired mechanism.

\begin{figure}[htbp]
\centering
\includegraphics[width=1\textwidth]{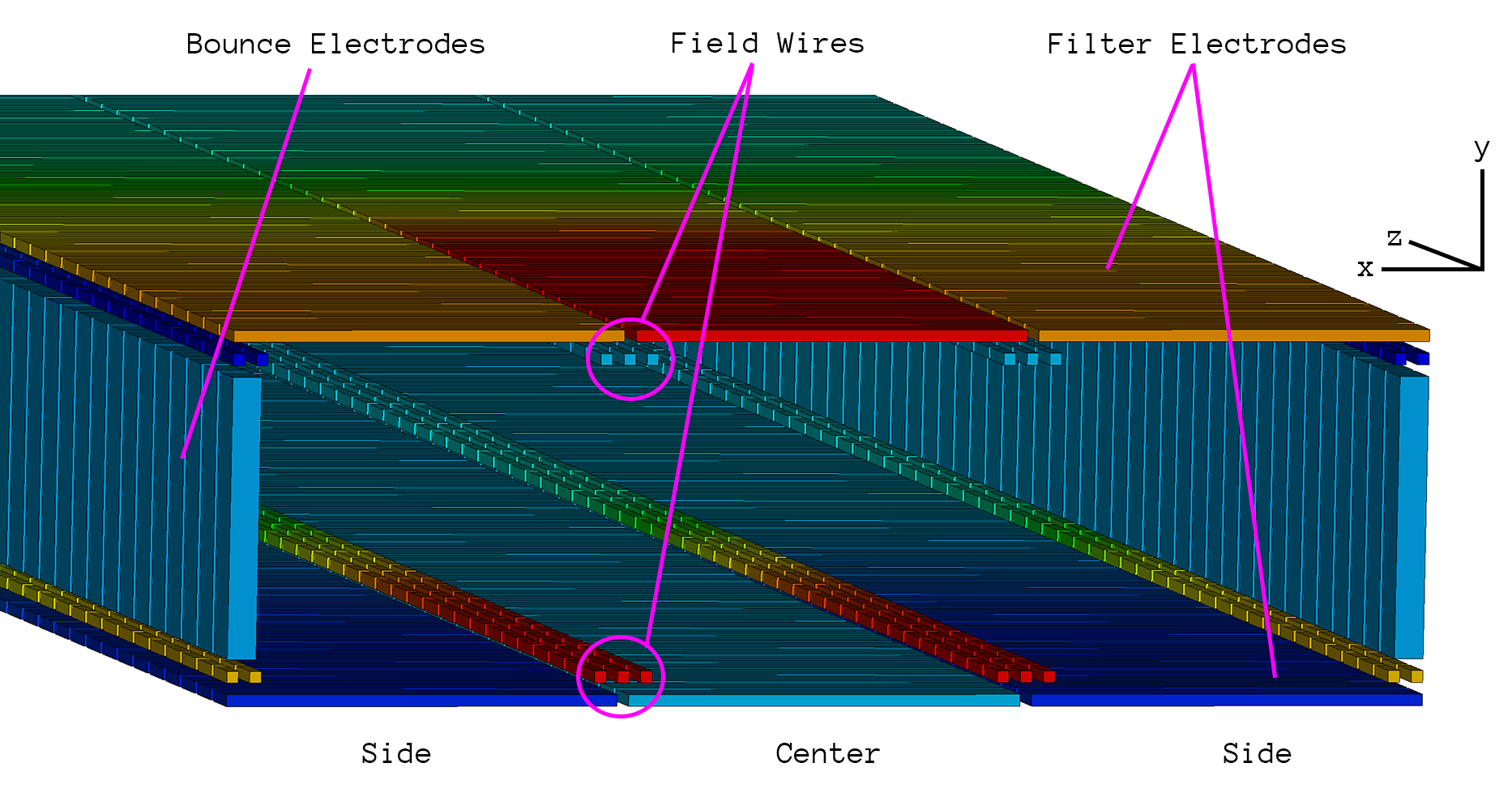}
\caption{Transverse perspective view of the three-channel filter geometry.
}
\label{fig:transview_pers}
\end{figure}

Explicitly, with $T_\perp^0$ and $T_\parallel^0$ the initial transverse and parallel kinetic energies, respectively, the idealized potentials are now, setting the reference offset $\phi_0 = T_{total}$,

\begin{align}
\label{eq:side_potentials}
    \phi_{cent}(z)|_{y=0}   & =  T_\perp^0\,e^{-z/\lambda_{cent}} + T_\parallel^0\,e^{-z/\lambda_{side}}  \\
    \phi_{side}(z)|_{y=0}   & =  \phi_{cent}(z) -T_\parallel^0\,e^{-z/\lambda_{side}} \nonumber \\
                            & =  T_\perp^0\,e^{-z/\lambda_{cent}} \ ,
\end{align}

where $\lambda_{side}$, which sets the rate of parallel energy drain, is not required to be the same as $\lambda_{cent}$. If $\lambda_{side} = \lambda_{cent}$,

\begin{align}
\label{eq:side_potentials_equal}
    \phi_{cent}(z)|_{y=0}  & =  T_{total}\,e^{-z/\lambda} \\
    \phi_{side}(z)|_{y=0}  & =  T_\perp^0\,e^{-z/\lambda} \ ,
\end{align}

which makes clear that the center potential is the total energy drained and the side potential is just the potential for an electron with the same transverse kinetic energy but zero parallel component, i.e. transverse-only draining. The corresponding electrode voltages are

\begin{align}
\label{eq:3ch_filtervoltages1}
    V_{cent,\,y_-}(z)   & = 0 \\
    V_{cent,\,y_+}(z)   & = 2\,T_{total}\,e^{-z/\lambda} \nonumber \\
                        & = 2\,T_\perp^0\,e^{-z/\lambda_{cent}} + 2\,T_\parallel^0\,e^{-z/\lambda_{side}} \\
    V_{side,\,y_-}(z)   & = 0 -2\,T_\parallel^0\,e^{-z/\lambda_{side}} \\
    V_{side,\,y_+}(z)   & = 2\,T_\perp^0\,e^{-z/\lambda_{cent}} + 2\,T_\parallel^0\,e^{-z/\lambda_{side}} \ .
\label{eq:3ch_filtervoltages2}
\end{align}

In contrast to the center, the addition of the parallel energy term to the side voltages is split across the top and bottom electrodes rather than applied only to the top electrodes. The total potential difference across the top and bottom in the side well is also increased relative to the center in order to increase the $E_y$ magnitude and therefore the forward drift in $z$ to account for the slight backwards motion in $z$ due to the curvature of the field lines.

\begin{figure}[htbp]
\centering
\includegraphics[width=1\textwidth]{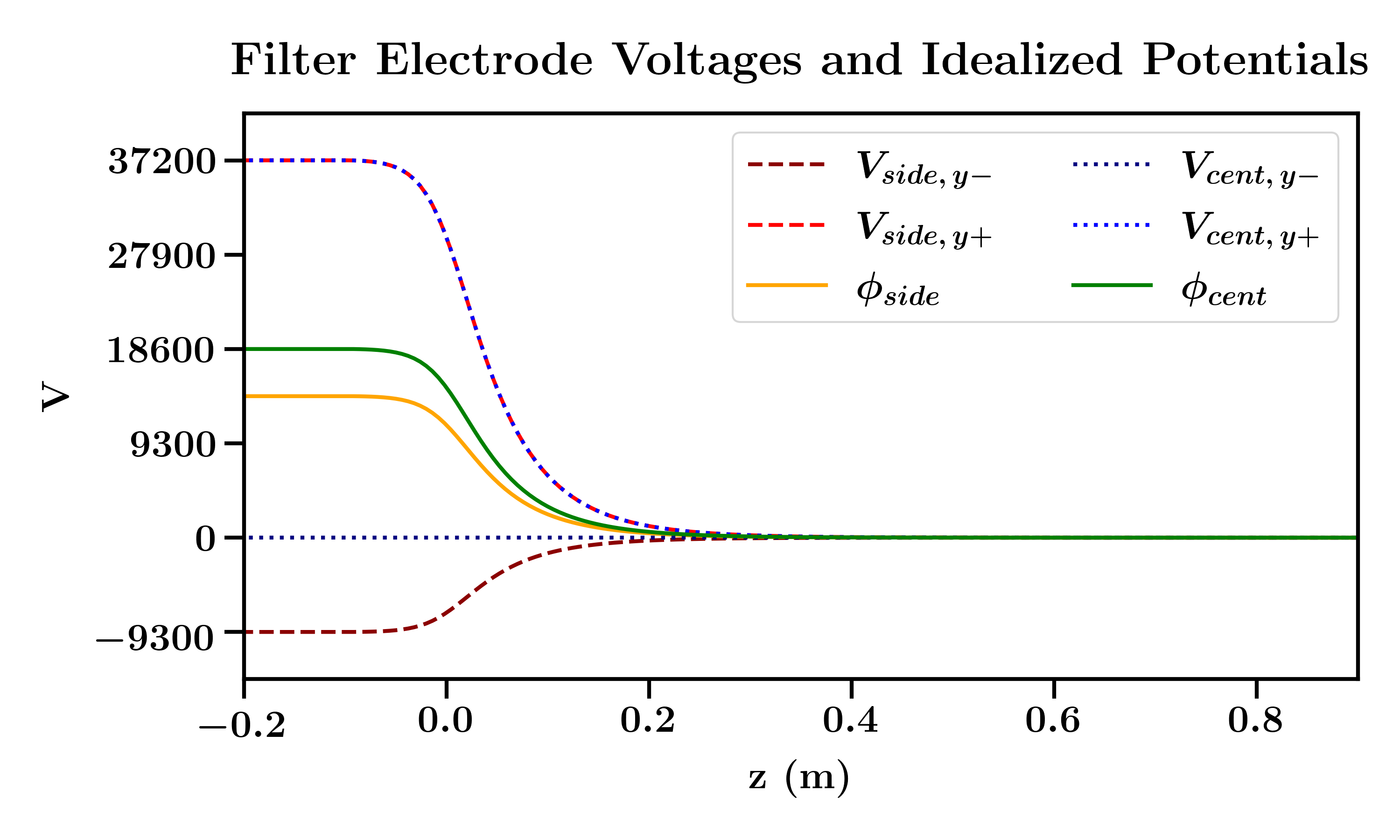}
\caption{Filter electrode voltages and idealized potentials along the center line of the respective wells. Voltages are set for a pitch 60$^\circ$ electron with total kinetic energy 18.6\,keV, with $\lambda_{side} = \lambda_{cent}$.
}
\label{fig:3ch_voltages_idealized_potentials}
\end{figure}

Given~\eqref{eq:3ch_filtervoltages1}-\eqref{eq:3ch_filtervoltages2}, the voltages are iterated as before, with the correction term also split across top and bottom for the side voltages.

\begin{align}
    V_{cent,\,y_-}[i] & = 0 \\
    V_{cent,\,y_+}[i] & = V_{cent,\,y_+}[i-1] + 2 \left( \phi_{ideal,\,cent} - \phi_{cent}[i-1]\right) \\
    V_{side,\,y_\pm}[i] & = V_{side,\,y_\pm}[i-1] + \left( \phi_{ideal,\,side} - \phi_{side}[i-1]\right)
\label{eq:3ch_correction}
\end{align}

\begin{figure}[htbp]
\centering
\includegraphics[width=1\textwidth]{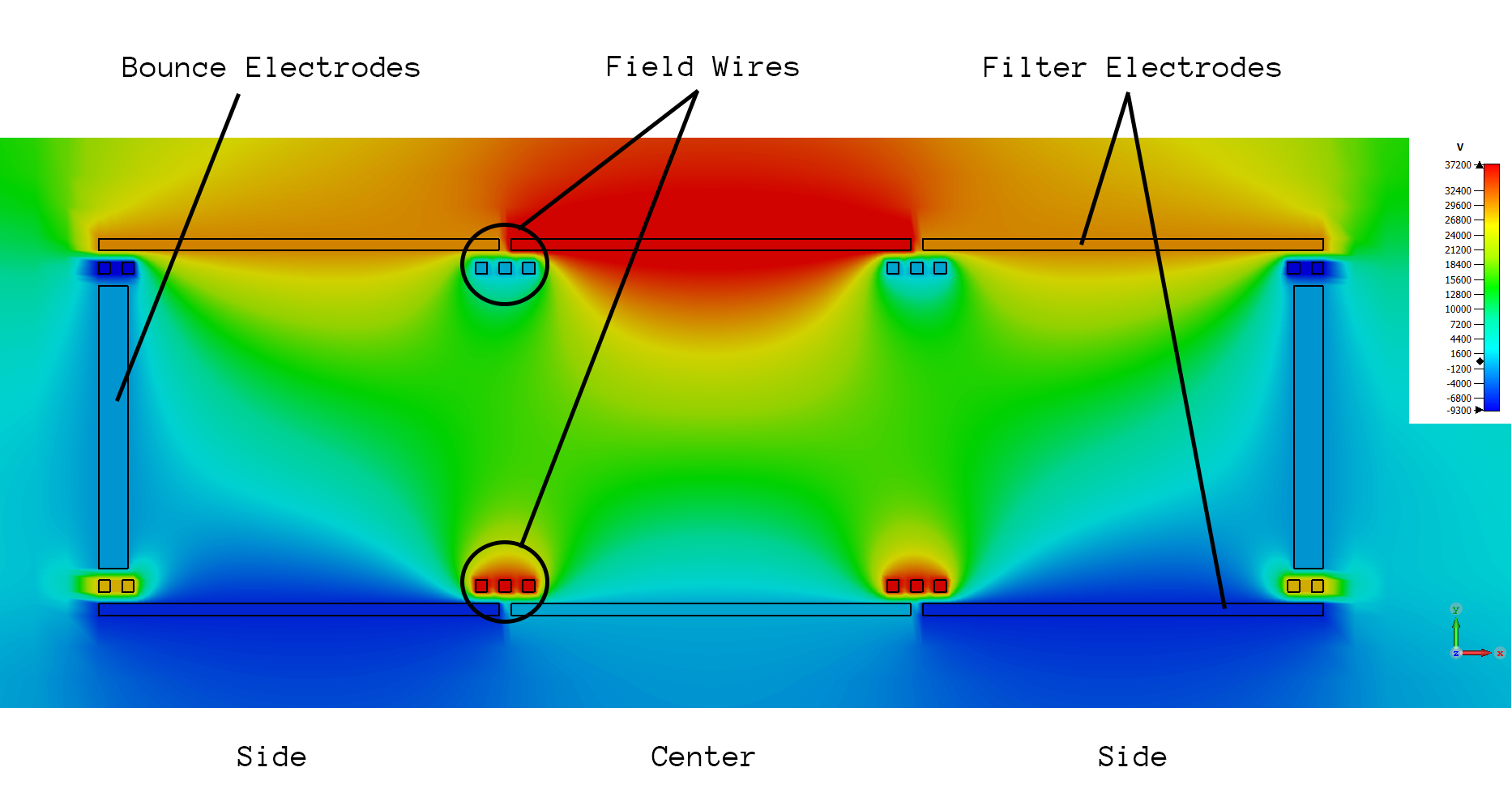}
\includegraphics[width=1\textwidth]{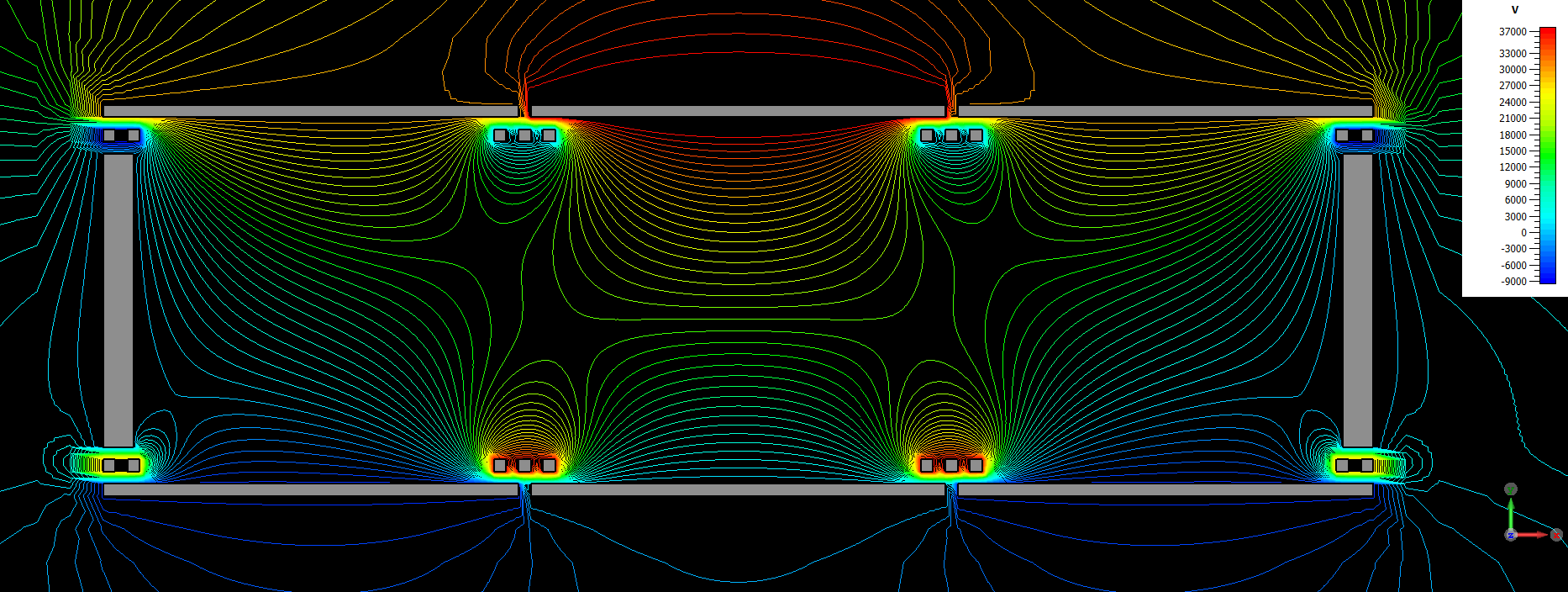}
\caption{Transverse cross-section of the three-channel filter geometry in the $xy$-plane at the start of the filter, with potential field overlay. Annotations (top) show the placement of the electrodes and field wires. The field wire voltages mirror the opposite side filter electrode voltages. Equipotential lines (bottom) show the sharpness of the transition between the side and center wells.
}
\label{fig:transview}
\end{figure}

\begin{figure}[htbp]
\centering
\includegraphics[width=1\textwidth]{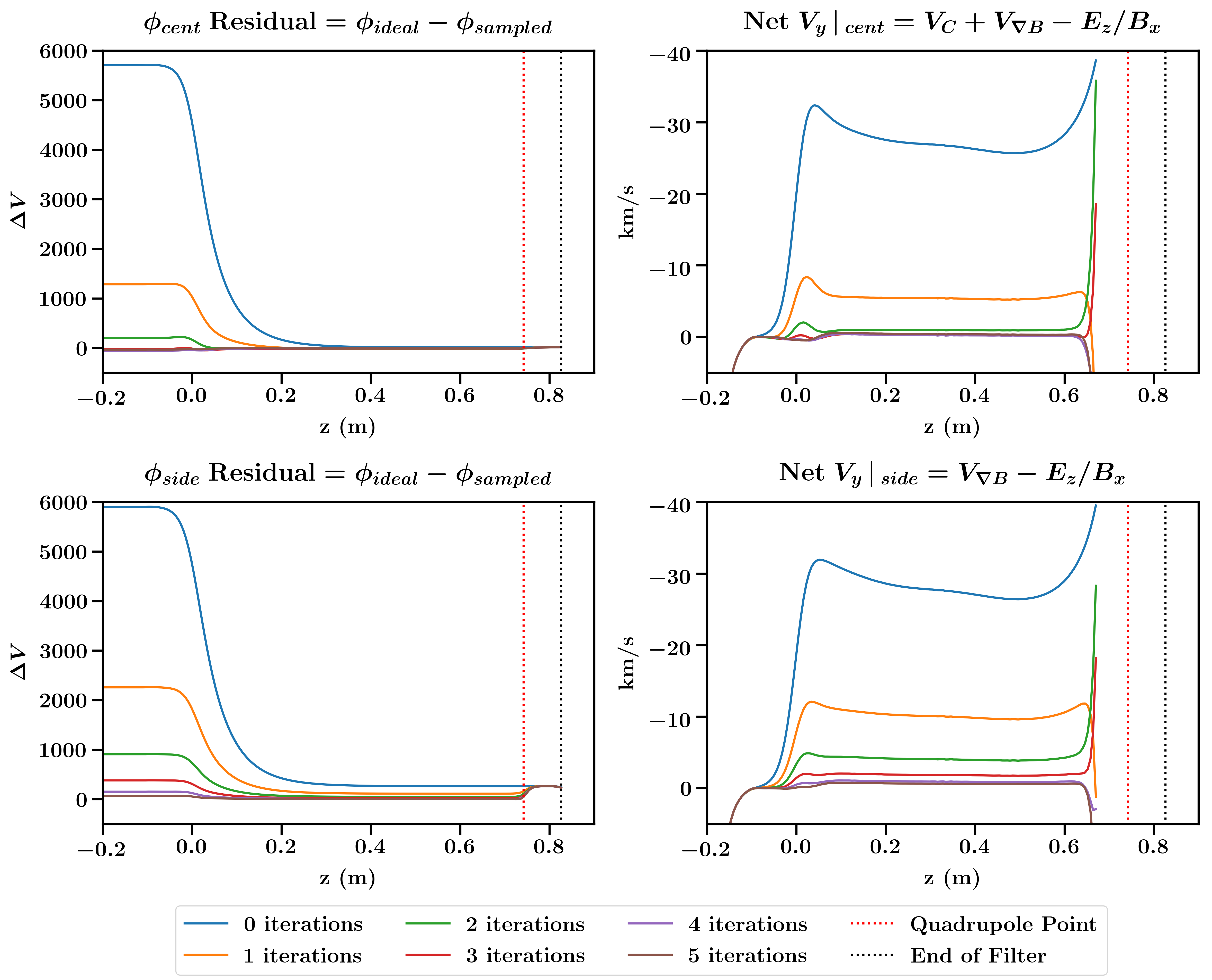}
\caption{Voltage iteration results for the three-channel filter geometry. Top row plots are for the center channel, bottom row plots for the side channels. The curvature drift $V_C$ for the center channel is calculated by assuming the parallel kinetic energy decreases as $T_\parallel^0 \, e^{-z/\lambda_{side}}$ in Eq.~\eqref{eq:gradbcurv}.
}
\label{fig:parallel_drain}
\end{figure}

\begin{figure}[htbp]
\centering
\includegraphics[width=1\textwidth]{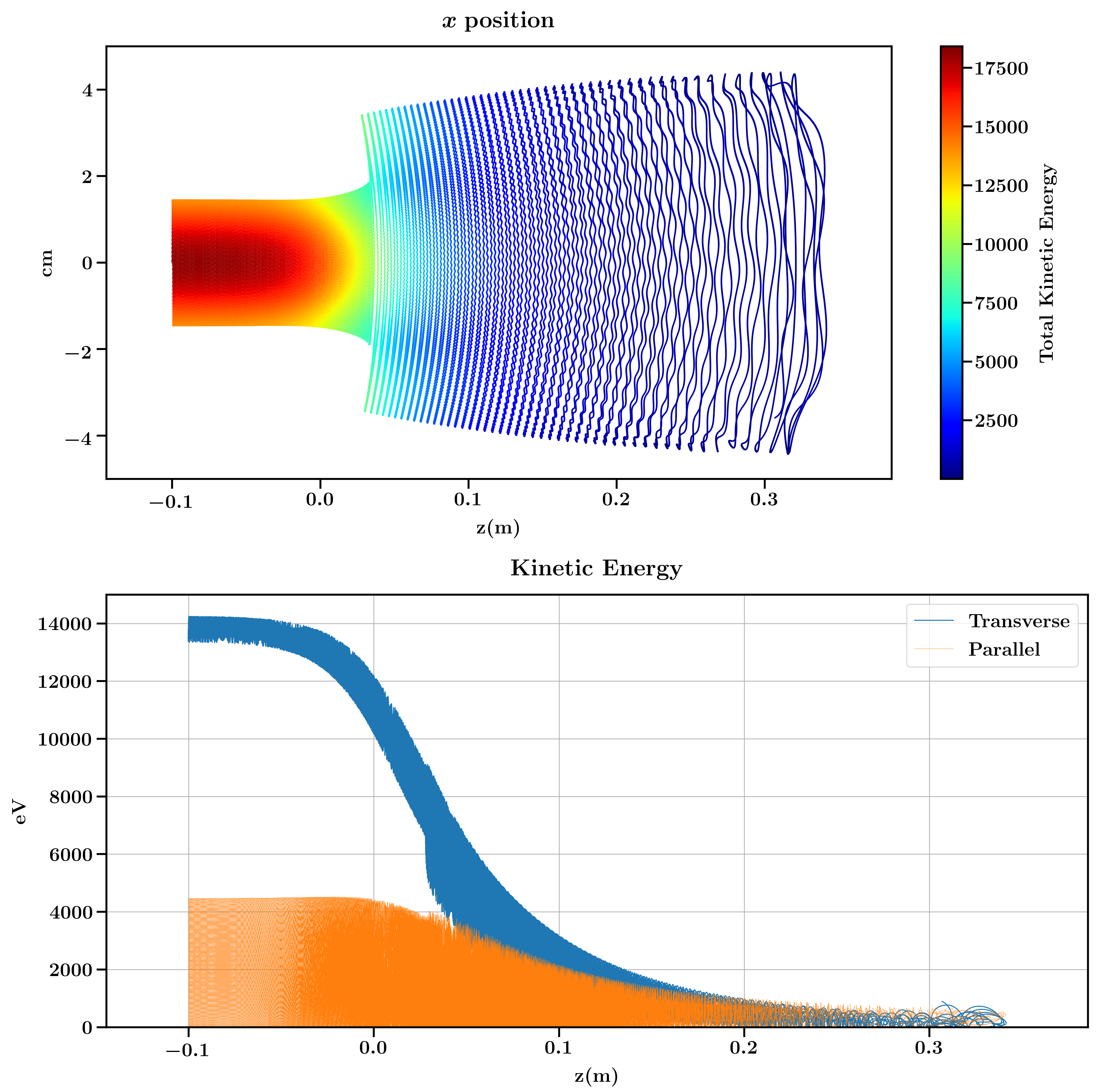}
\caption{$x$-position and kinetic energy drain for a pitch 60$^\circ$ electron in the first 35\,cm of the three-channel filter showing the simultaneous draining of parallel and transverse kinetic energies.
}
\label{fig:p60_x_ke}
\end{figure}

Finally, to increase the sharpness of the transition between the center and side wells, additional field wires are placed along the splits in the filter electrodes. To maintain maximum relative uniformity of the side and center potentials down the entire filter length $z$, both the field wires and bounce electrodes can be segmented in $z$ in the same fashion as the filter electrodes. In practice this level of discrete granularity in the field wires is not necessary along the entire filter length if the aspect ratio is small enough. The field wire voltages mirror the opposite side filter electrode voltages and the bounce electrodes, to ensure continuous bouncing but not so much so that the bounce potential contaminates the side and center potentials, are set with a fixed offset plus a decay term proportional to the initial parallel kinetic energy,

\begin{align}
    V_{wires,\,y_\pm}(z)    & = V_{cent/side,\,y_\mp}(z) \\
    V_{bounce}(z)   & = \phi_0 - V_{fixed} - \alpha T_\parallel^0\,e^{-z/\lambda} \ .
\label{eq:bounce_voltages}
\end{align}

where $\alpha < 1$. A transverse cross-section of the three-channel filter geometry is shown in Figure~\ref{fig:transview}. Figure~\ref{fig:transview_pers} shows a perspective view. For $\lambda_{side} = \lambda_{cent}$, results of the voltage iteration are shown in Figure~\ref{fig:parallel_drain} for a pitch 60$^\circ$ electron. Figure~\ref{fig:p60_x_ke} shows the x-position parallel to the field lines and the simultaneous draining of transverse and parallel kinetic energies with the three-well scheme.

\section{Filter Entry and Exit}

In this section we briefly illustrate schemes for electron transport into and out of the filter. A detailed analysis of the parameters and acceptance criteria for end-to-end transport are the subject of an upcoming study and not described here.

\subsection{Einzel Lens at Exit}

At the end of the filter, the electron kinetic energy has been drained down to a few eV and the magnetic field is around one milliTesla.  To extract the electron for final readout, the cyclotron motion must be unraveled into a linear trajectory leading to the calorimeter. To unravel the cyclotron motion, a cylindrical $\mu$-metal housing is interfaced to the end of the filter, inside of which the magnetic component of the Lorentz force law is zero. To direct the resulting trajectory to the calorimeter, inside the $\mu$-metal housing is an Einzel lens, i.e. a series of three conducting rings with the two outer rings set at a higher potential than the middle ring. The lens acts as a pure electrostatic focusing device analogous to a convex optical lens, focusing charged particle trajectories instead of light.

\begin{figure}[htbp]
\centering
\includegraphics[width=1\textwidth]{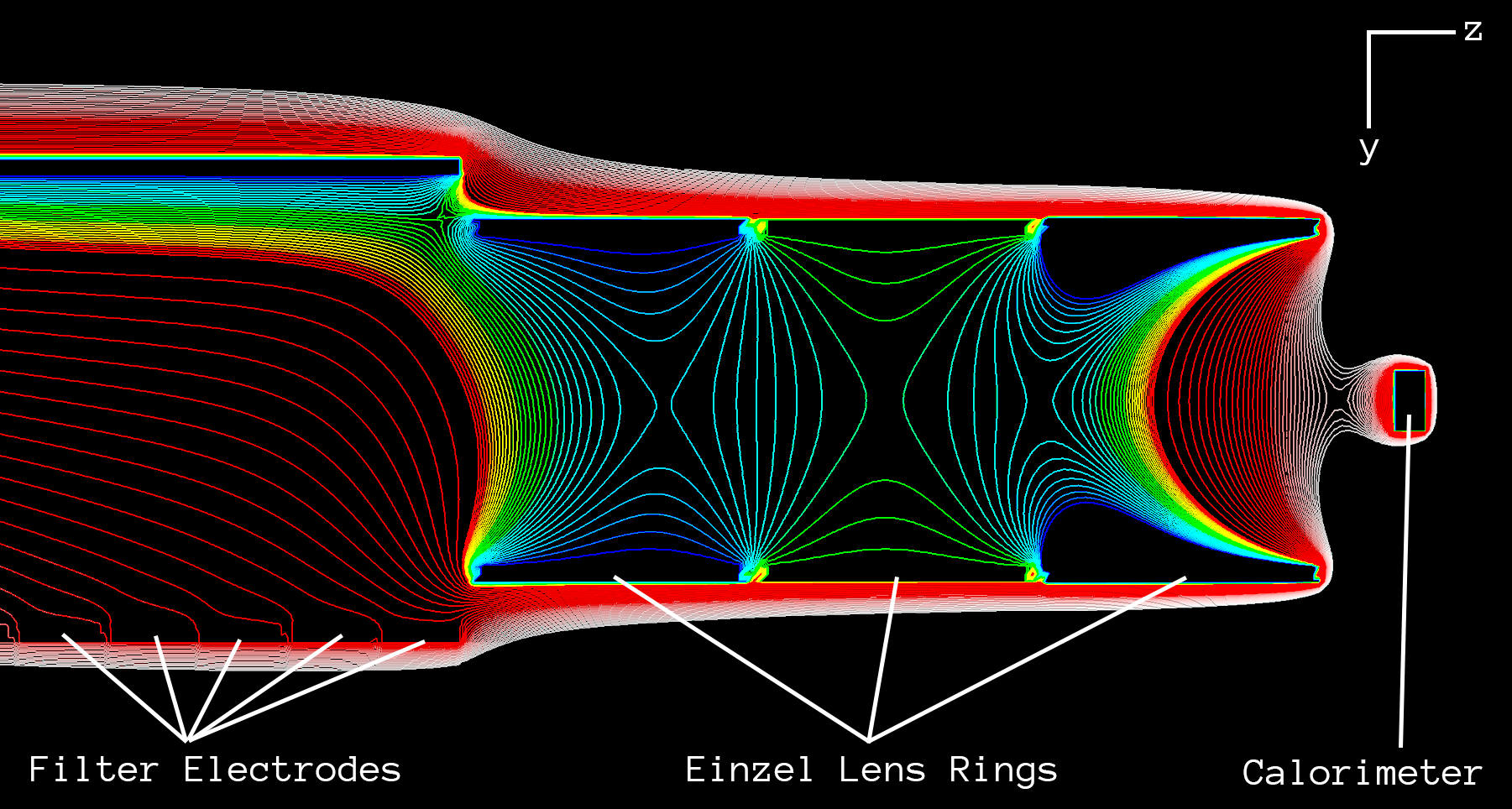}
\includegraphics[width=1\textwidth]{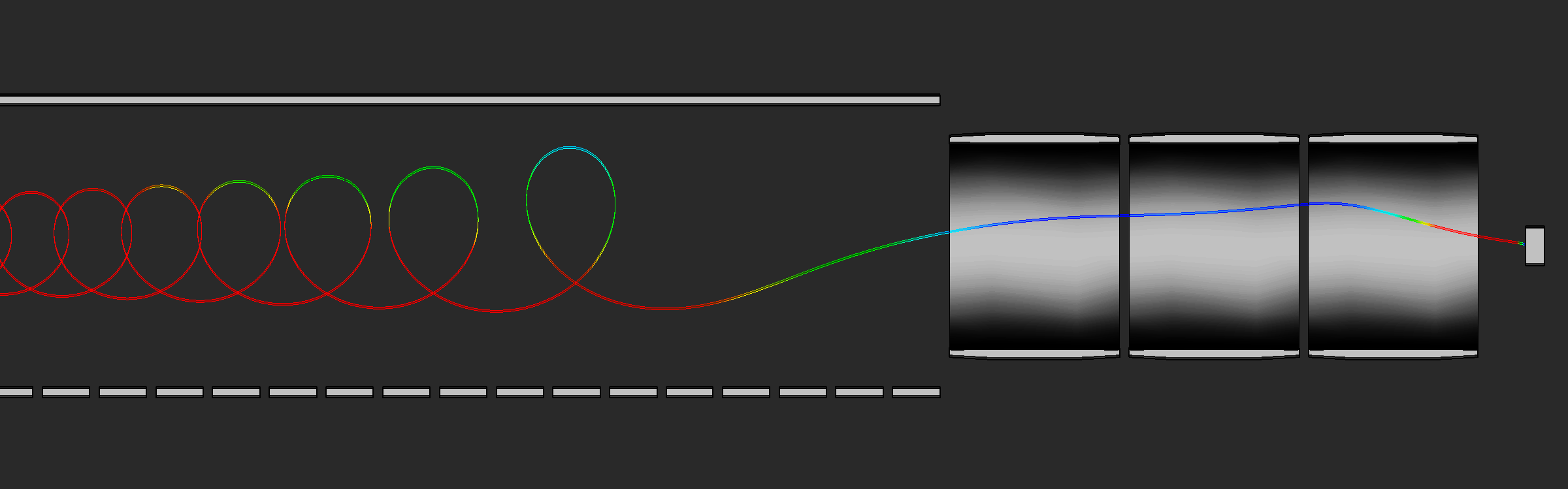}
\caption{ (\textit{top}) Equipotential lines in the plane $x=0$ at the end of the filter showing the Einzel lens and microcalorimeter interface. (\textit{bottom}) Sample trajectory of an electron at the end of the filter. The cyclotron motion is unraveled and focused through the Einzel lens to the calorimeter. The $\mu$-metal housing, not pictured, is concentric with the Einzel lens.
}
\label{fig:einzel_lens}
\end{figure}

The calorimeter is located in the focal plane of the lens on the exit side of the filter. As the final electron kinetic energy is uncertain at the level of 1-10~eV depending on the filter starting field and RF momentum measurements, there is a spread in final electron trajectories distributed over the acceptance of the microcalorimeter. Equipotential lines and a sample trajectory demonstrating the maneuver are shown in Figure~\ref{fig:einzel_lens}.

\subsection{Parallel Momentum Drain on Entry}

\begin{figure}[htbp]
\centering
\includegraphics[width=1\textwidth]{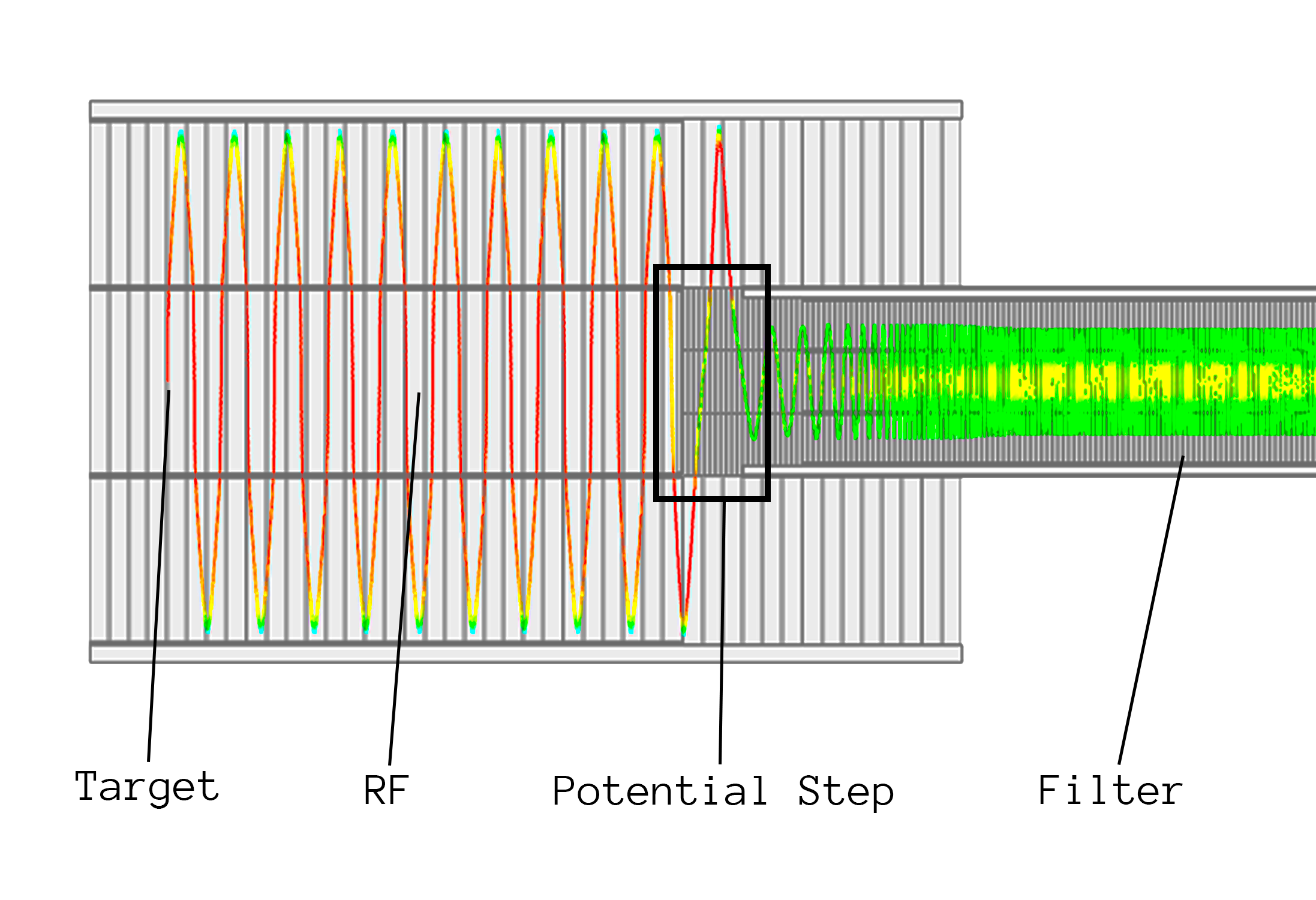}
\caption{Entry into the filter from the RF. Trajectory color indicates kinetic energy from high (red) to low (green). Drift speed is exaggerated for clarity. From left to right, an electron is released from the target source. The RF module gets an initial estimate of the parallel and transverse momentum splits, which is used to dynamically set the filter voltages. The filter protrudes into the RF to pick up the electron via a potential step that drains most of the parallel energy on entry.
}
\label{fig:concept_endtoend}
\end{figure}

On the entry side, the electron is at full kinetic energy and, as described in~\cite{betti2019design}, estimates of the parallel and transverse momentum component splits are given by the Radio-Frequency (RF) antenna region.  The RF region is larger than the filter and the filter juts into the RF as shown in Figure~\ref{fig:concept_endtoend}. The reference potential of the filter, $\phi_0$, is set above the reference potential of the RF by an amount approximately equal to the parallel kinetic energy, creating a potential step. When an electron crosses the potential step of the filter while undergoing bounce motion in the RF, its parallel momentum is reduced to almost zero while the transverse momentum is largely unchanged. The electron is then captured into the filter by forward $\bm{E \times B}$ drift before it leaves the potential step. The scheme is similar to the three-channel filter in that a potential step is used to drain the parallel momentum while moving the electron in the transverse direction. This time the voltage of the central region, the filter, is set above the sides, which are a part of the RF.

Although the three-channel filter shows that it is possible to drain both parallel and transverse energy components at the same time within the filter, by far the most efficient way to drain the parallel energy is to remove a large portion of it in one step upon entry. Since the transverse energy is largely unchanged, such a maneuver effectively increases the pitch angle of the electron inside the filter. The three-channel scheme can then be used to drain the remaining parallel energy. An example trajectory of this entry mechanism is shown in Figure~\ref{fig:concept_endtoend}.

\subsection{Injection of Charged Particles with a Reversed Filter}

The compact size and dynamic setting of the electrode voltages invites an additional perspective on running the filter "in reverse" as an entry mechanism into the filter proper, and separately as a standalone particle accelerator. A reversed filter starts with a charged particle at low kinetic energy and has a reversed direction of gradient-$B$ with mirrored filter voltages.  With the same drift balancing conditions, this accelerates the charged particle's transverse kinetic energy along the center line rather than drain it, hence a reversed filter is a new type of particle accelerator.  There is no linear acceleration in this process.  The electron's magnetic orbital angular momentum is accelerated during a process of constant drift into high magnetic field. The ability of a reversed filter to inject charged particles into a magnetic field is potentially useful in applications such as plasma heating and lithography.

\begin{figure}[htbp]
\centering
\includegraphics[width=1\textwidth]{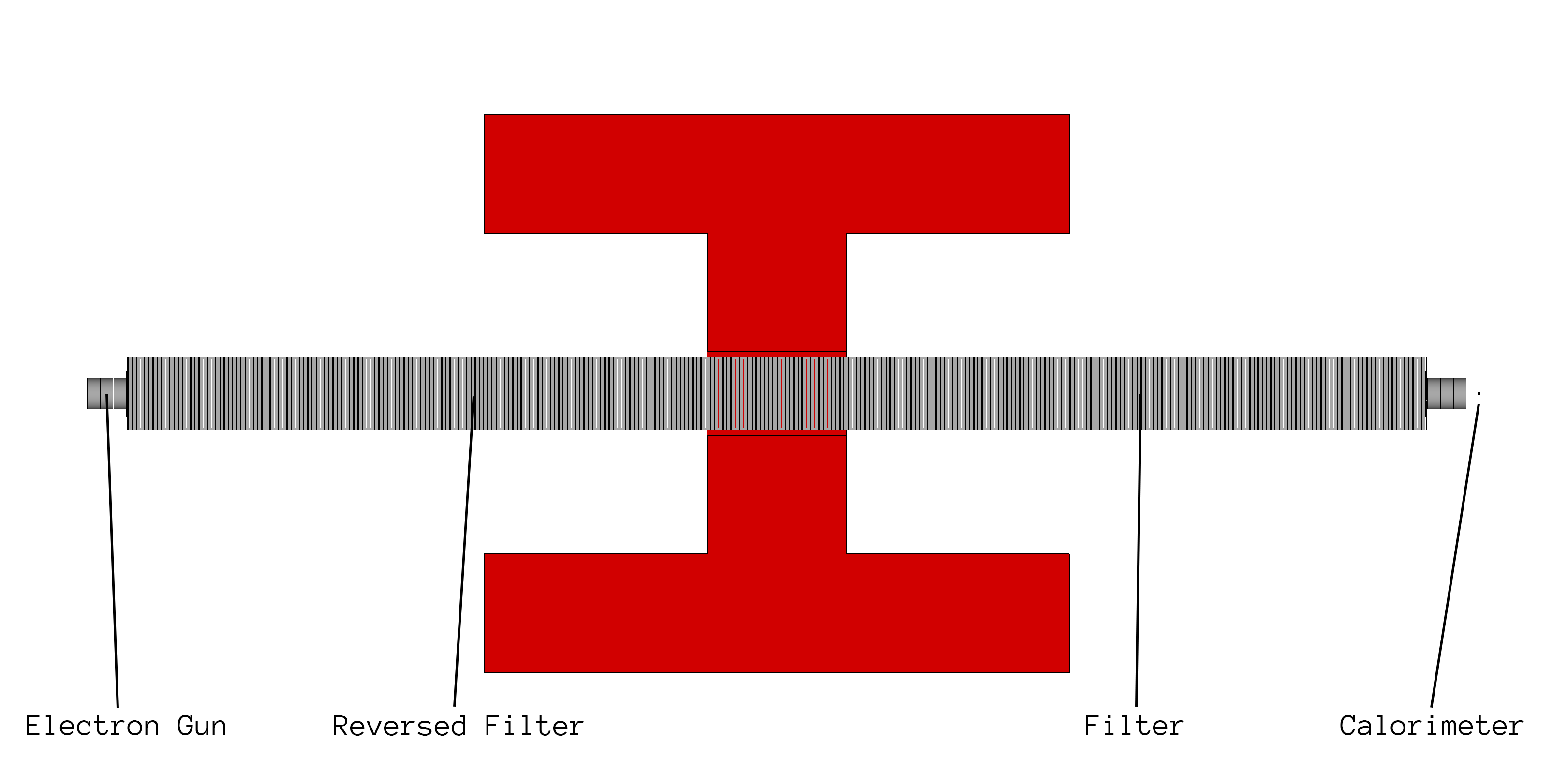}
\includegraphics[width=1\textwidth]{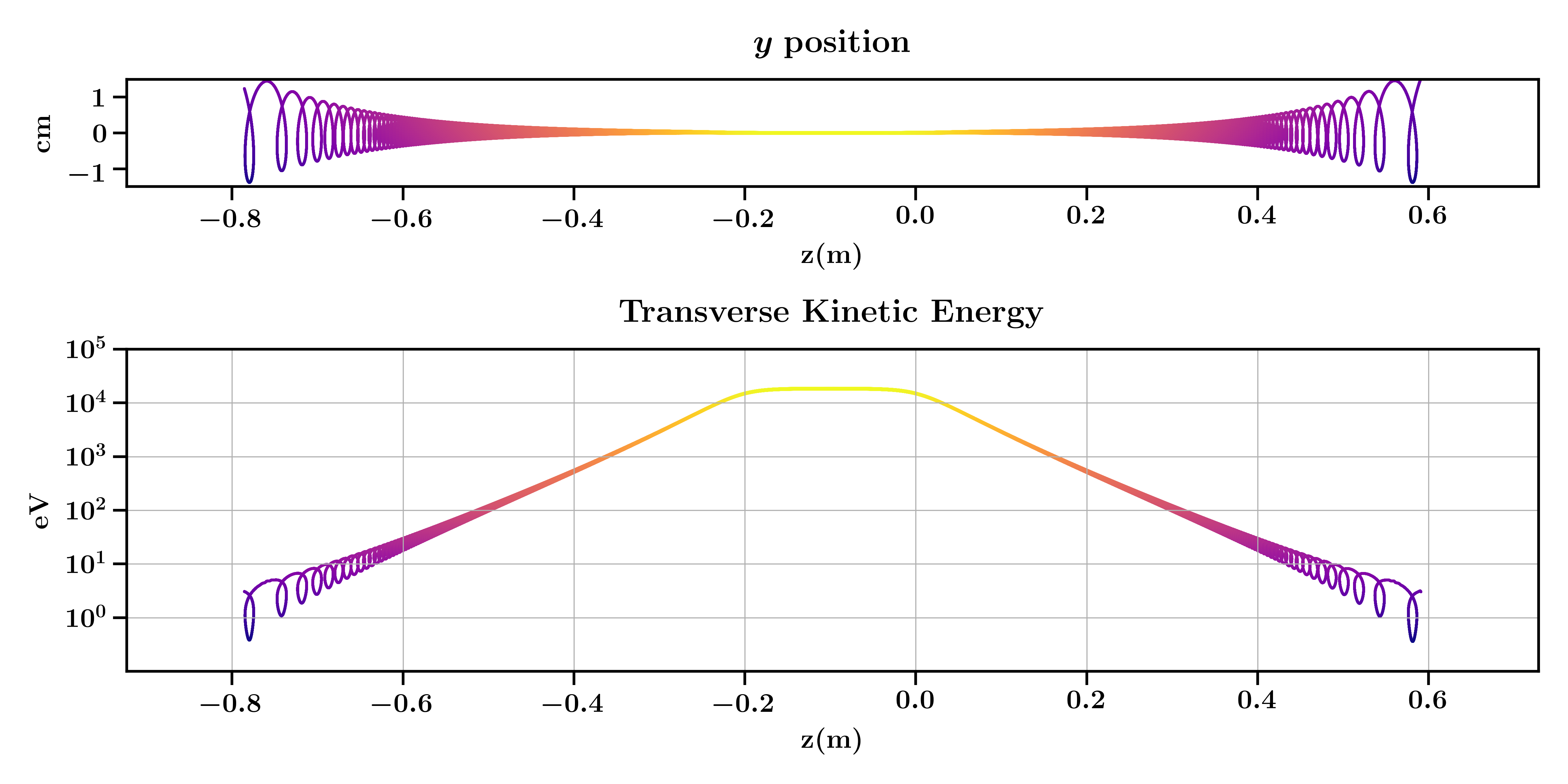}
\caption{(\textit{top}) Schematic of a reverse filter setup. Both magnet and filter are mirrored about the center of the magnet. Injected electrons on the left are accelerated by gradient-$B$ drift using the same drift conditions as the draining process. (\textit{bottom}) Trajectory and transverse kinetic energy for a pitch 90$^\circ$ electron in the reverse filter setup. The electron starts at $\approx$ 1\,eV, is accelerated by gradient-$B$ drift to $\approx$ 18.6\,keV, then drained back down to $\approx$ 1\,eV. Highlighted overlays are GCS values.
}
\label{fig:reversed_filter}
\end{figure}

A reversed filter can be realized with the current design by mirroring the magnet extensions and filter electrodes about the center of the magnet (Figure~\ref{fig:reversed_filter}). In this setup an electron is emitted from an electron gun at around 1\,eV, accelerated into the uniform $B$ region to 18.6\,keV, then filtered back through the filter down to 1\,eV again.

For the PTOLEMY experimental program, a reversed filter provides a new tool to study transverse drift filter operation without the complexities of interfacing with the RF, but it does not replace the need to benchmark the filter performance with an isotropically emitted electron conversion line source, as is typical for tritium endpoint spectrometers~\cite{sentkerestiova2017gaseous}.  The combination of isotropic source data and acceptance calculations and transmission efficiency data from the reverse filter may be another approach to identifying systematic inefficiencies in the transverse filter performance.


\section{Construction of PTOLEMY demonstrator}

\begin{figure}[htbp]
\centering
\includegraphics[width=1\textwidth]{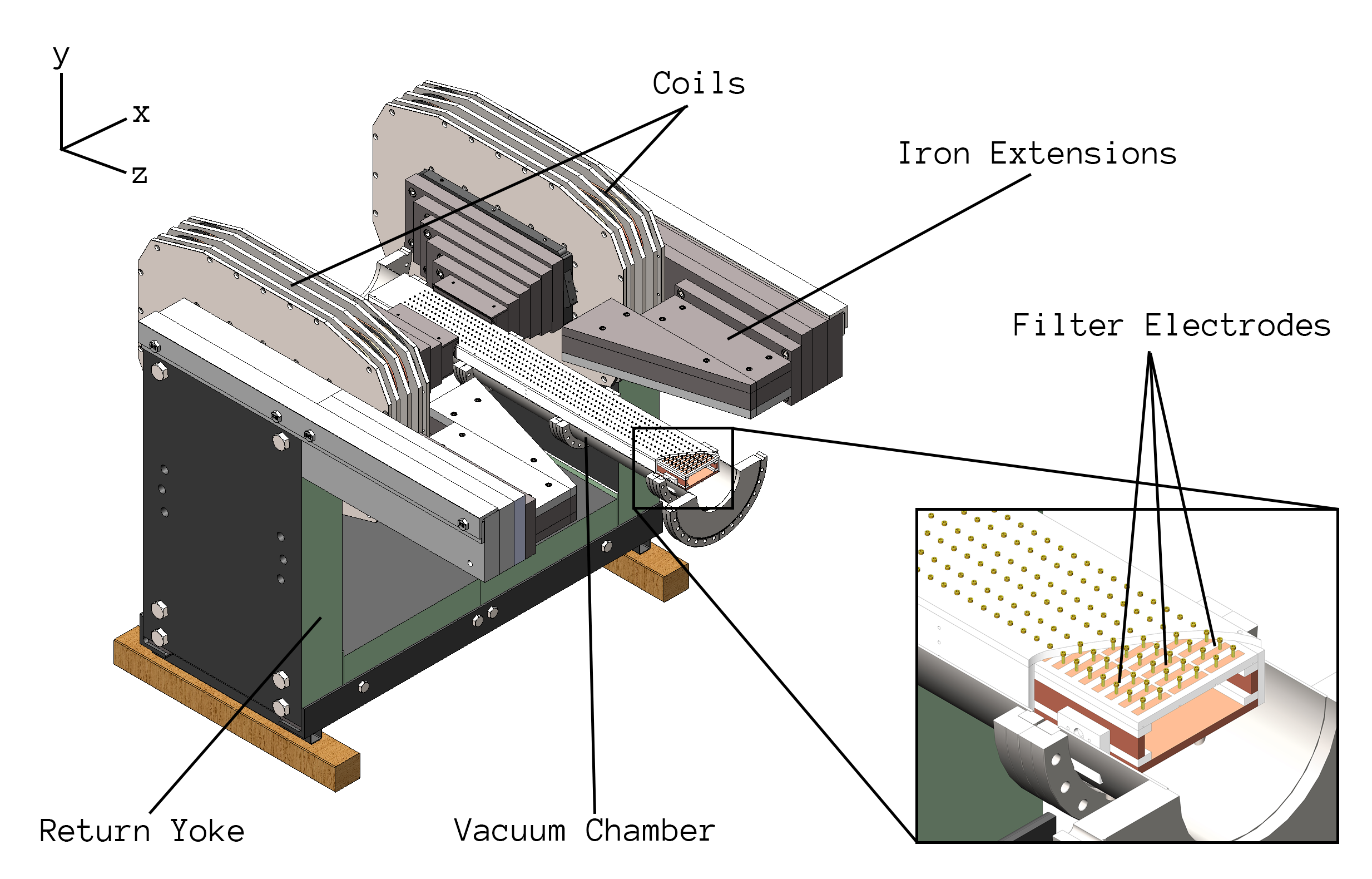}
\caption{Mechanical concept image of the PTOLEMY demonstrator implementing the iron-extension magnet and three-channel filter design for the top electrodes. A single bottom electrode is used in this prototype image. Trapezoidal iron add-ons are used on the extensions for fine shaping of the field. 
}
\label{fig:filter_mechanical}
\end{figure}

\begin{figure}[htbp]
\centering
\begin{minipage}[c][11cm][t]{.64\textwidth}
  \vspace*{\fill}
  \centering
  \includegraphics[width=\linewidth]{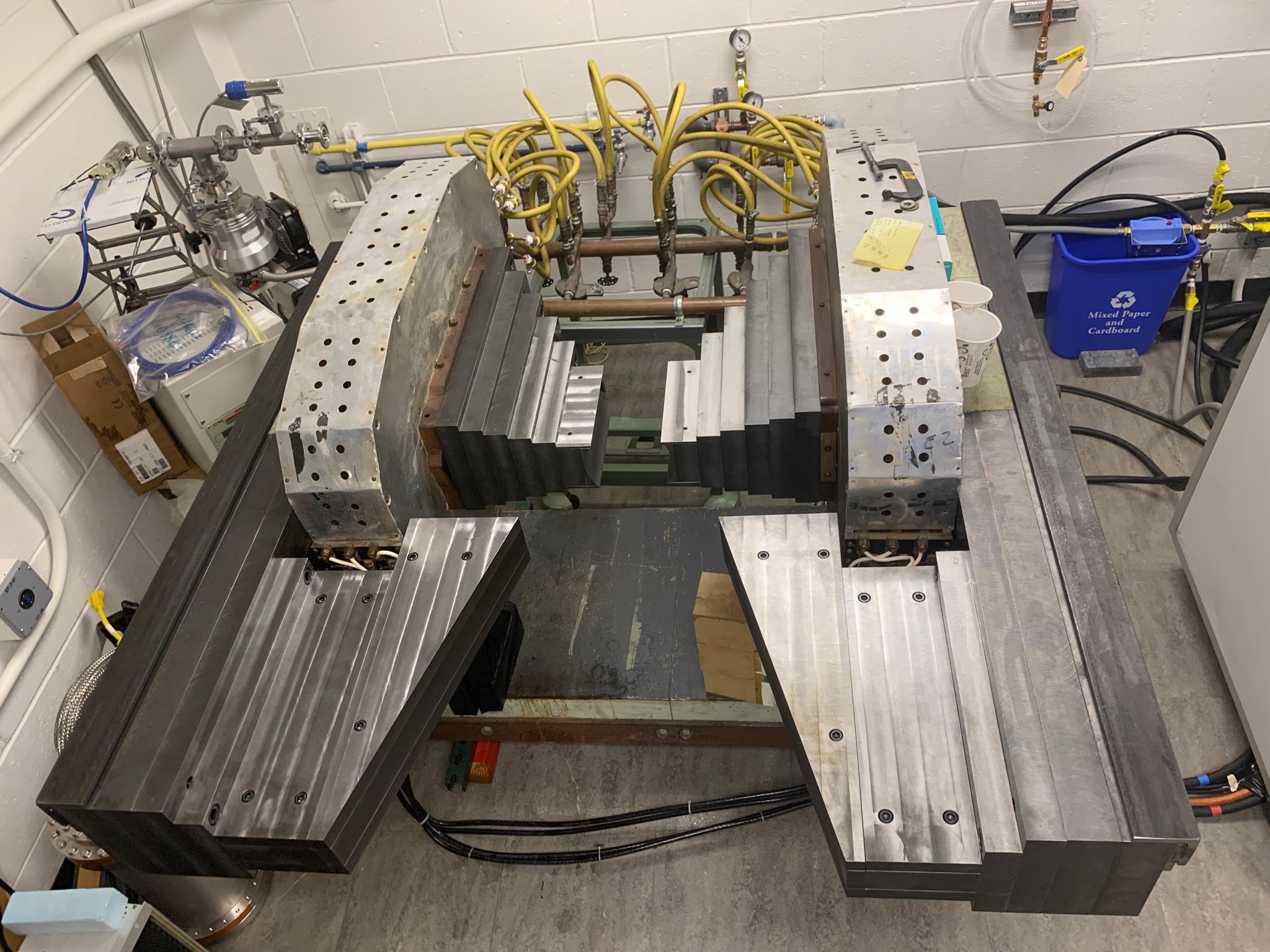}
  \label{fig:test1}
\end{minipage}%
\begin{minipage}[c][11cm][t]{.32\textwidth}
  \vspace*{\fill}
  \centering
  \includegraphics[width=\linewidth]{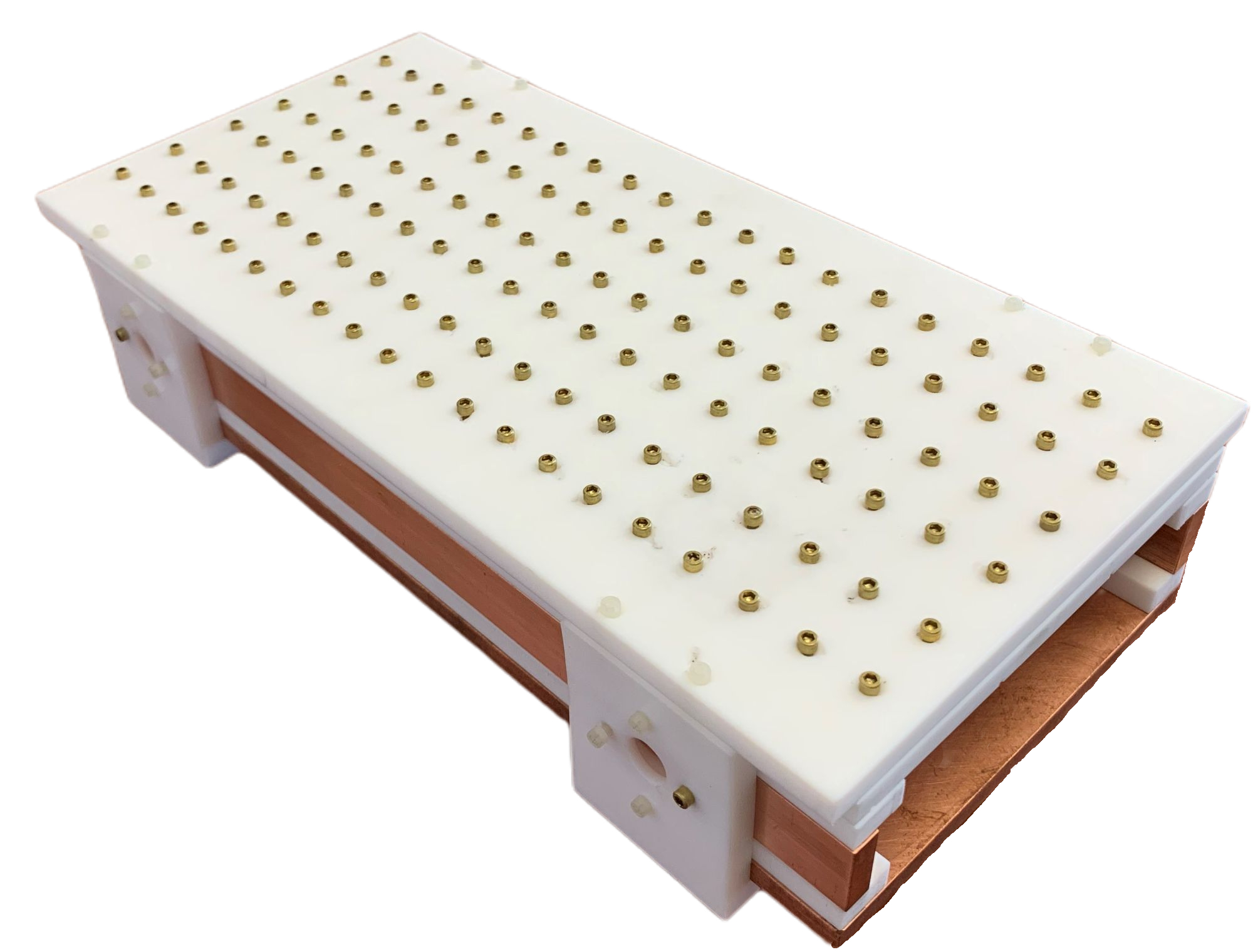}
  \label{fig:test2}
  \includegraphics[width=\linewidth]{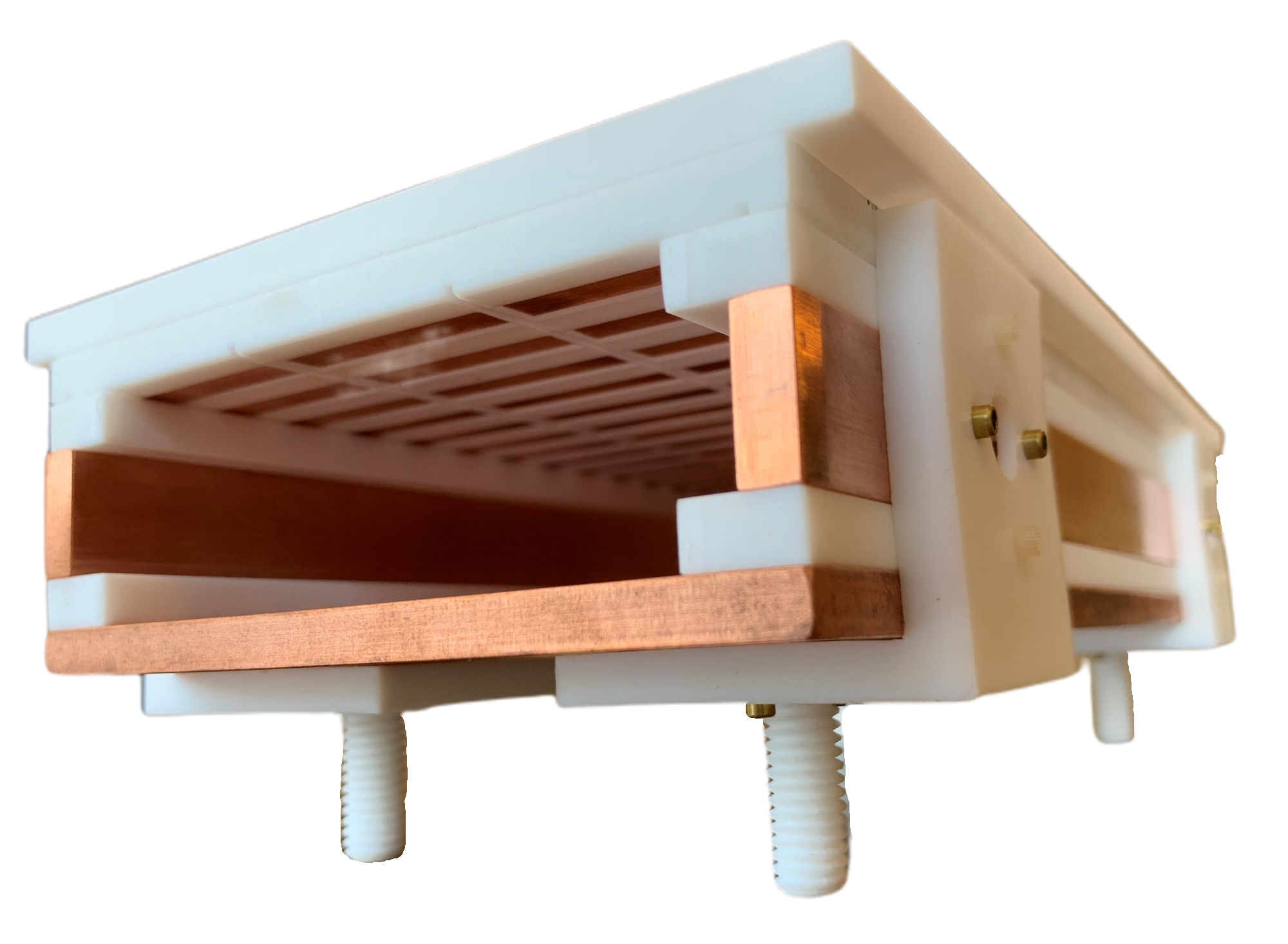}
  \label{fig:test3}
\end{minipage}
\caption{(\textit{left}) Photo of iron-extension design magnet at Jadwin Hall, Princeton University, Princeton, NJ. (\textit{right}) Photo of a short version of the three-channel filter with a single-piece bottom electrode.
}
\label{fig:jadwin_magnet}
\end{figure}

Construction of a PTOLEMY demonstrator implementing the iron-extension magnet and three-channel filter designs is underway. Mechanical designs are shown in Figure~\ref{fig:filter_mechanical}. Figure~\ref{fig:jadwin_magnet} shows the assembled magnet at Princeton University, Princeton, NJ.

\begin{figure}[htbp]
\centering
\includegraphics[width=0.45\textwidth]{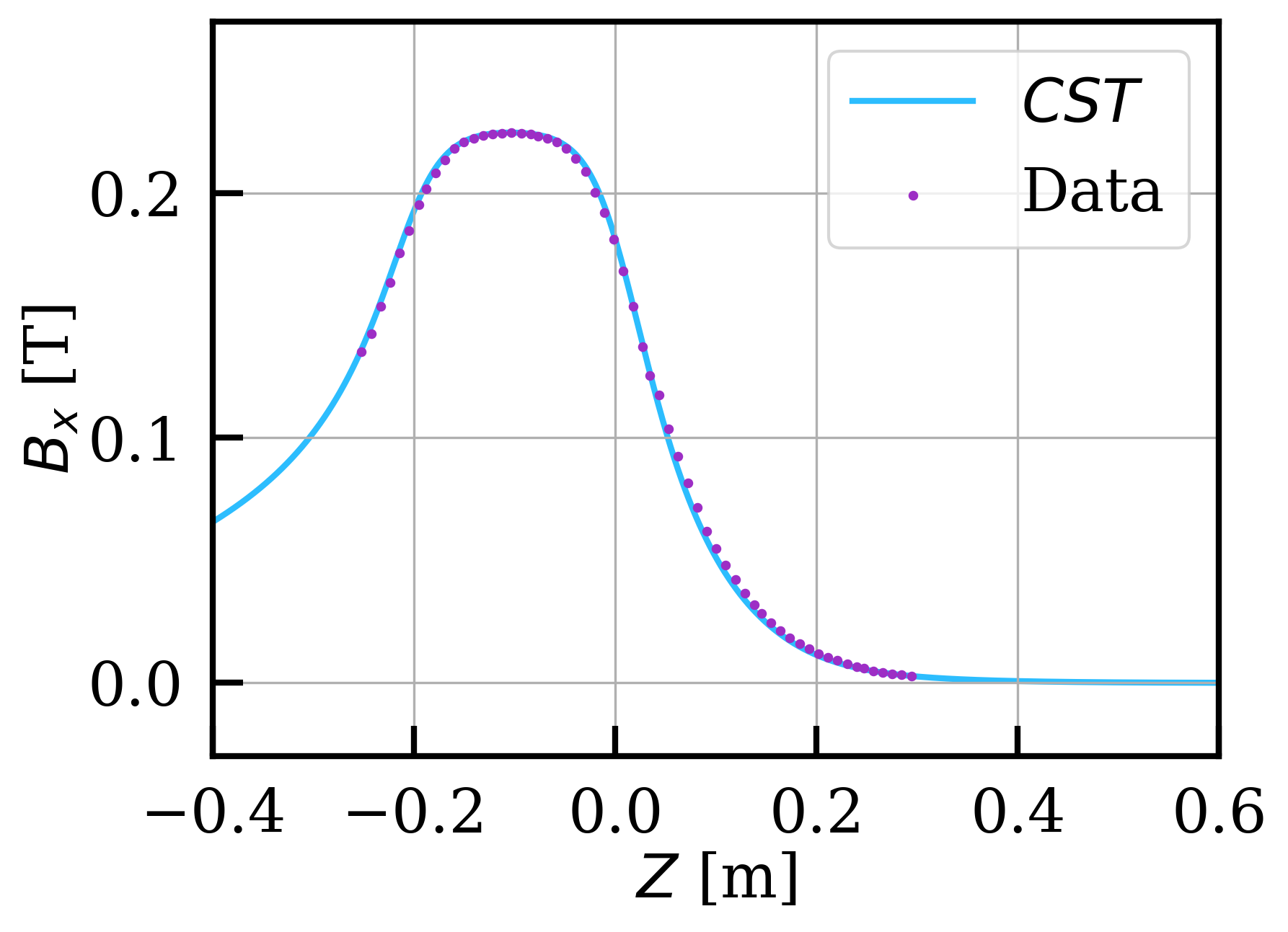}
\includegraphics[width=0.45\textwidth]{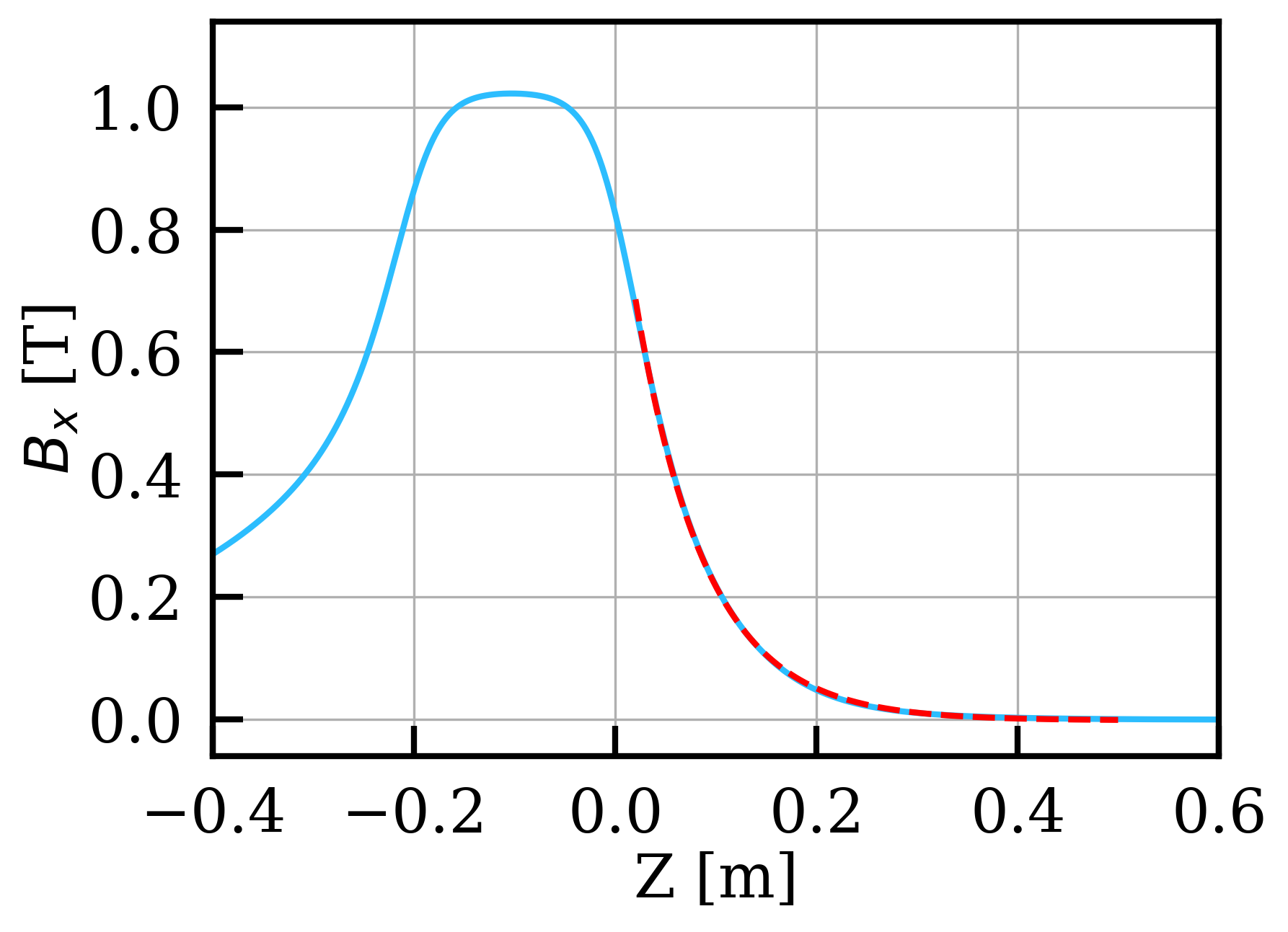}
\caption{(\textit{left}) $B_{x}$ with modest current power supplies along the z-axis vs. $B_{x}$ in simulation using the CST software suite. (\textit{right}) $B_{x}$ in simulation using new high power supplies.  The plateau region between the pole faces is about 1 T. The red dashed line is the fitting curve with an exponential and the characteristic length is 6.98~cm.
}
\label{fig:demo_Bx}
\end{figure}

The $B_{x}$ field was mapped out by digital 3-axis hall magnetic sensors.
Initial investigation at low power found good agreement between the measured and simulated fields as shown in Figure~\ref{fig:demo_Bx} (\textit{left}). At high power with new power supplies, the $B_{x}$ field will start at about 1~T between the pole faces and fall exponentially with $\lambda\sim 7$~cm as shown in Figure~\ref{fig:demo_Bx} (\textit{right}).

\subsection{Tolerance Estimation}
\label{subsec:tolerance}

\begin{figure}[htbp]
\centering
\includegraphics[width=0.45\textwidth]{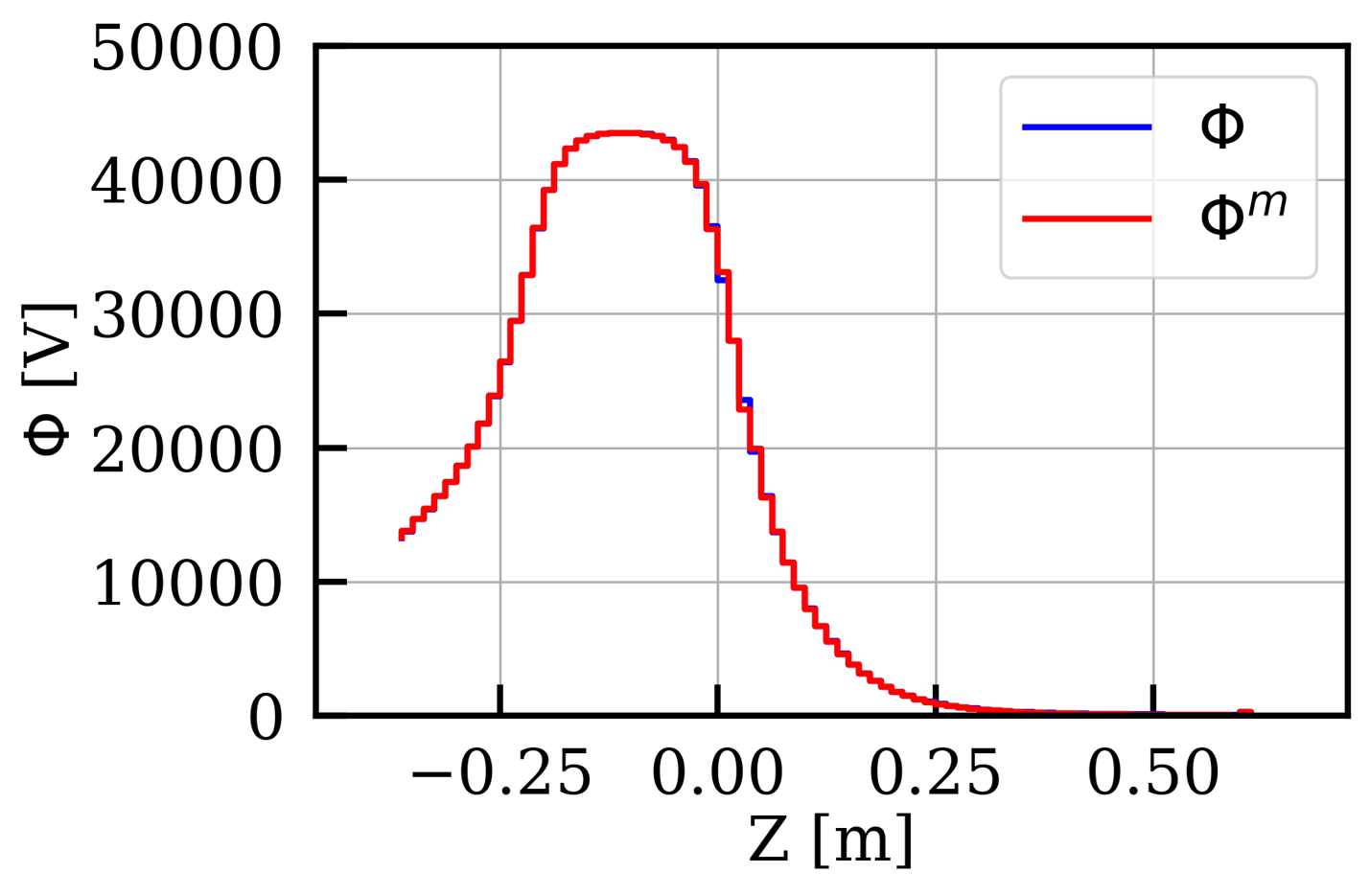}
\includegraphics[width=0.45\textwidth]{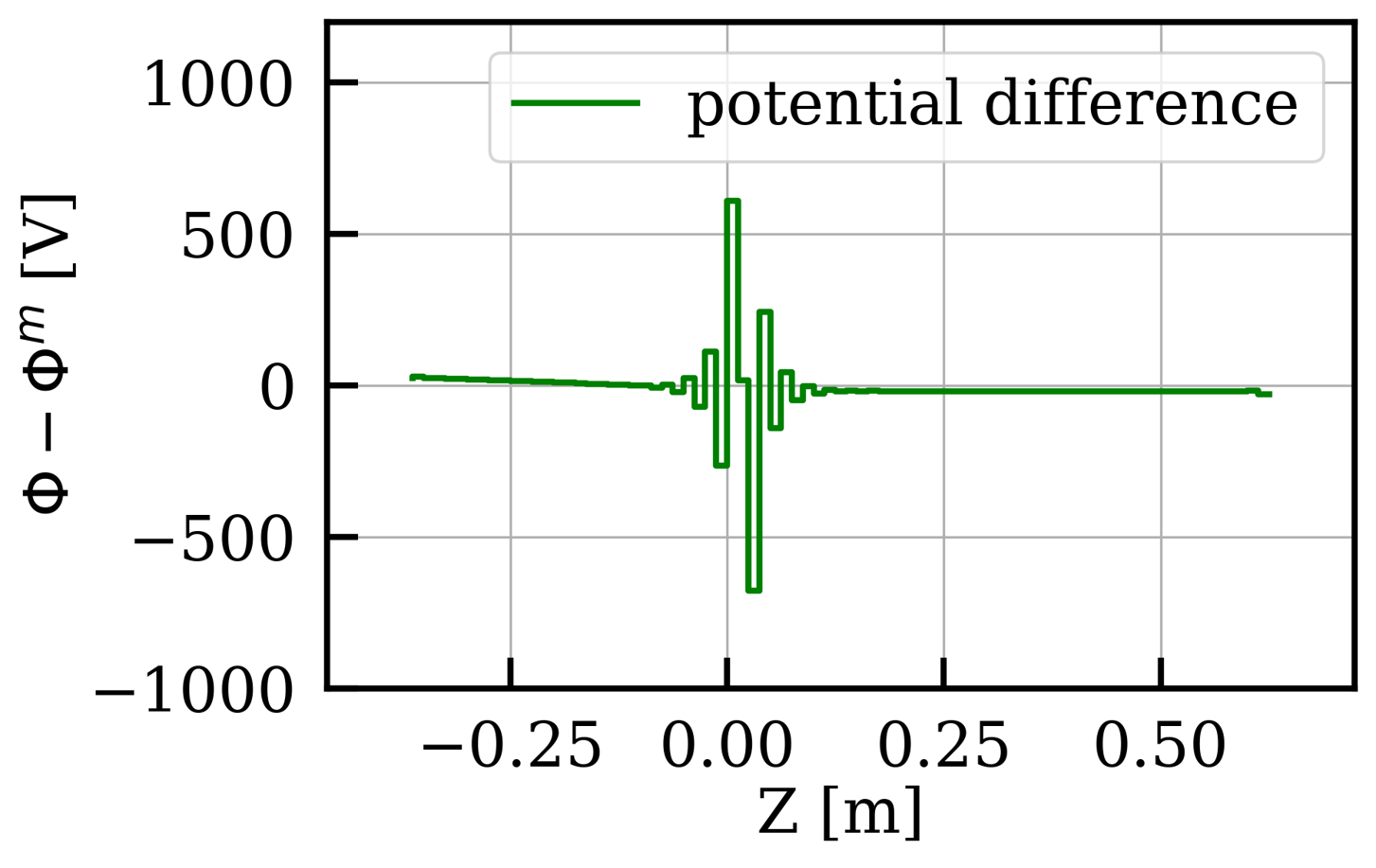}
\caption{(\textit{left}) Voltage profiles on the $+y$ electrodes computed by the boundary value method for the original geometry~(blue) vs. a geometry with a displaced electrode~(red); (\textit{right}) The difference between the original and corrected voltages.
}
\label{fig:filter_voltage_diff}
\end{figure}

The filter electrodes shown in Figure~\ref{fig:filter_mechanical} have 80 individual voltages set over a distance of one meter along the $z$ direction.  Each electrode has a width of 6.25\,mm and a separation gap of 6.25\,mm between closest surfaces. Using the simulated $B$ field in Figure~\ref{fig:demo_Bx}, the voltage settings are computed using the foregoing boundary-value method, as shown in Figure~\ref{fig:filter_voltage_diff}.

\begin{figure}[htbp]
\centering
\includegraphics[width=0.7\textwidth]{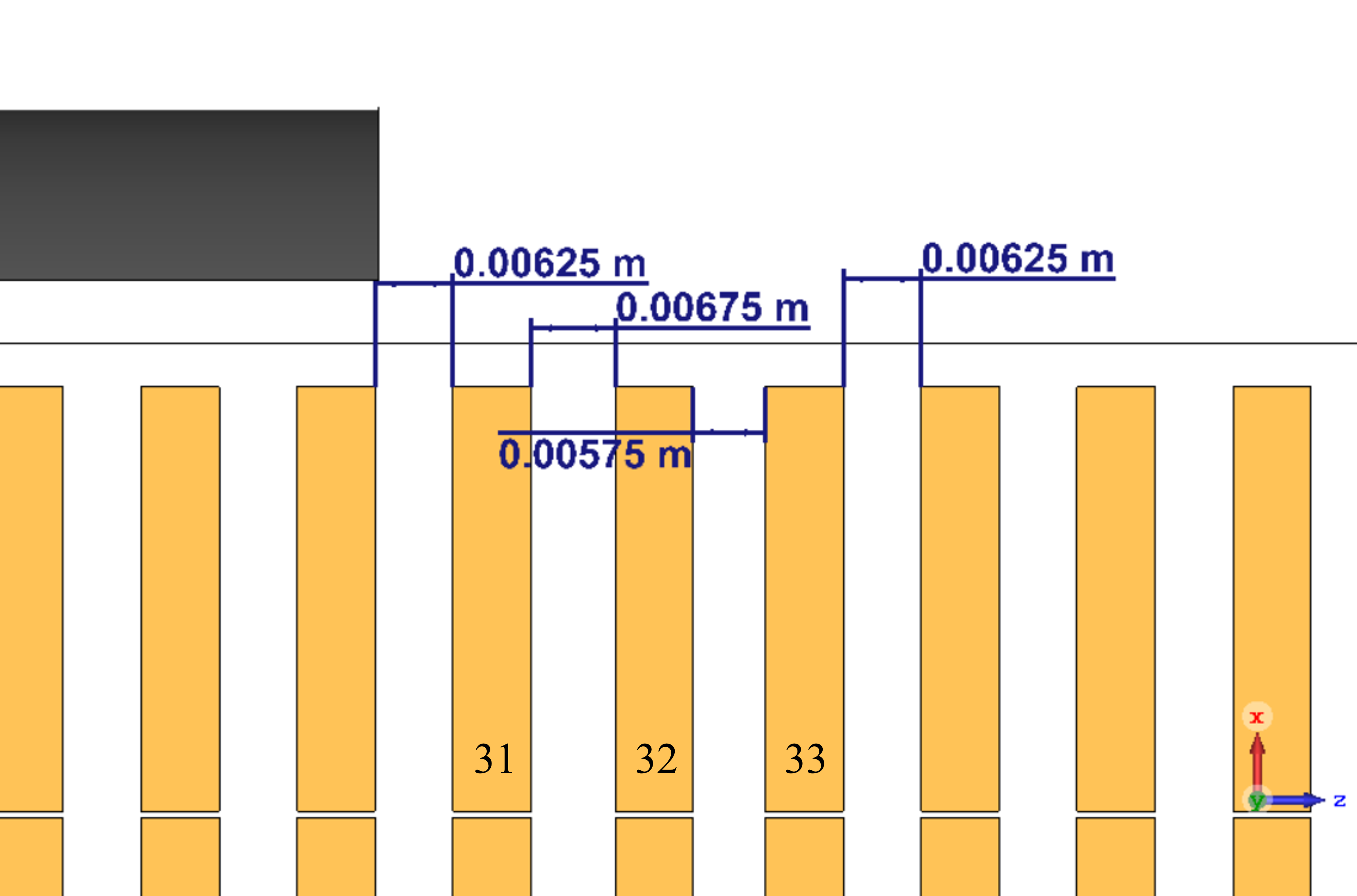}
\caption{Detailed view of the $+y$ electrodes around the displaced 32$^\textrm{nd}$ electrode. Black numbers indicate electrode numbers. The blue annotations measure the distance between adjacent electrodes.
}
\label{fig:filter_displaced}
\end{figure}

To accommodate tolerance from machining and installation, a displacement was added to the 32$^\textrm{nd}$ electrode in  simulation as shown in Figure~\ref{fig:filter_displaced}, where the magnitude of the $B$ field gradient is largest.

\begin{figure}[htbp]
\centering
\includegraphics[width=
\textwidth]{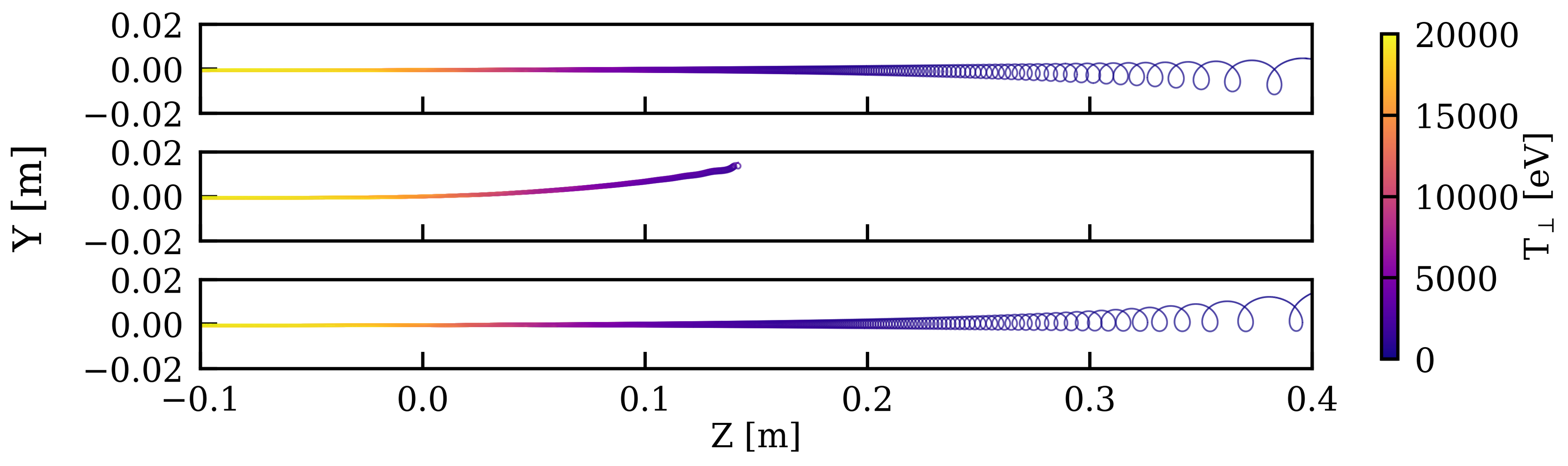}
\caption{Electron trajectories with the original geometry and voltage setting~(top), with the displaced electrode and unchanged voltage setting~(middle), and with the displaced electrode and corrected voltage setting~(bottom).
}
\label{fig:filter_electron_traj}
\end{figure}

Applying the boundary-value method on the new geometry yields an updated set of voltages on the electrodes. As shown in Figure~\ref{fig:filter_electron_traj}, within 2\%, the corrected voltage settings on the 32$^\textrm{nd}$ and contiguous electrodes restore the balanced transverse drift to the required accuracy.

\section{Conclusion}

Implementation and optimization methods for the PTOLEMY transverse drift electromagnetic filter are studied. The design for an iron-core magnet that can produce the required field profile is presented. The idealized filter electrode voltages published in~\cite{betti2019design} are corrected to account for field transitions in a real-life setting; an iterative method and a boundary-value method are presented. A non-zero parallel kinetic energy introduces additional drift terms which must be mitigated by keeping the parallel component sub-dominant to the transverse. A new three-channel geometry was designed to implement a parallel draining mechanism. Schemes for entry into and exit from the filter are presented.  This report includes the first mechanical prototypes, construction designs and operating procedures for the PTOLEMY transverse drift filter.  The basis for the transverse drift filter described herein provides a foundation for a forthcoming end-to-end PTOLEMY concept paper for the direct detection of the Cosmic Neutrino Background.
\section*{Acknowledgments}
\addcontentsline{toc}{section}{Acknowledgments}

CGT is supported by the Simons Foundation (\#377485).

\bibliographystyle{JHEP}
\bibliography{main} 

\end{document}